\documentclass[lettersize,journal]{IEEEtran}

\usepackage{url,hyperref}
\usepackage{amsfonts}       
\usepackage{nicefrac}       
\usepackage{mathtools}      
\usepackage{color,bm}
\usepackage{mathrsfs}
\usepackage{amsmath}
\usepackage[linesnumbered,ruled,vlined]{algorithm2e}
\usepackage{graphicx} 
\usepackage{multirow}
\usepackage{amssymb}
\usepackage{mathtools}
\usepackage{amsthm}
\usepackage[skins]{tcolorbox}

\newtheorem{definition}{Definition}
\newtheorem{theorem}{Theorem}
\newtheorem{lemma}{Lemma}
\newtheorem{proposition}{Proposition}
\newtheorem{corollary}{Corollary}
\newtheorem{conjecture}{Conjecture}
\newtheorem{example}{Example}

\title{Bayesian Theory of Consciousness as Exchangeable Emotion-Cognition Inference}

\author{%
  Xin Li,\textsuperscript{1}\thanks{This work was partially supported by NSF IIS-2401748 and BCS-2401398. The author has used ChatGPT models to assist in the development of theoretical ideas presented in this paper.} \\
  Department of Computer Science\\
  University at Albany\\
  \texttt{xli48@albany.edu}
}
\begin{document}

\maketitle

\begin{abstract}
This paper presents a unified theoretical framework in which consciousness emerges as a cycle-consistent, affectively anchored inference process, recursively structured by the interaction between emotion and cognition. Drawing from information theory, optimal transport (OT), and the Bayesian brain hypothesis, we formalize emotion as a low-dimensional structural prior and cognition as a specificity-instantiating update mechanism. The resulting \emph{emotion–cognition cycle} minimizes joint uncertainty by using emotionally valenced states to guide cognitive selection, while cognitive appraisals reciprocally regulate emotional states. This recursive inference loop provides a computational bridge to the challenging problem of consciousness, reframing subjective experience as the informational footprint of temporally extended, affect-modulated simulation.
We extend this framework by introducing the \emph{Exchangeable Integration Theory of Consciousness} (EITC), in which temporally distributed emotional–cognitive episodes are modeled as conditionally exchangeable samples drawn from a latent affective self-model. This latent variable enables integration by supporting a coherent cause–and–effect structure with nonzero irreducibility, and differentiation by preserving the specificity of each episode under a shared generative prior. Crucially, we unify this structure with the Bayesian theory of consciousness through \emph{Rao-Blackwellized inference}, which marginalizes latent self-structure while enabling context-sensitive belief updates. This mechanism bridges exchangeability with stability, coherence, and goal-directed simulation, capturing both local adaptability and global consistency.
These dynamics are formalized via theorems grounded in De Finetti’s exchangeability principle, information integration theory, and KL-regularized optimal transport. Together, these contributions suggest that consciousness is not a static property but an evolving, distributed architecture, shaped by emotion, refined by cognition, stabilized by exchangeability, and unified through recursive inference over latent self-structure.

\end{abstract}


\section{Introduction}
\label{sec:1}

Emotion is evolutionarily older than cognition \cite{bennett2023brief}. Long before organisms developed the capacity for symbolic reasoning, planning, or abstraction, they evolved affective systems to regulate perception, behavior, and homeostasis under uncertainty \cite{panksepp1998affective}. These emotion systems encoded biologically salient constraints, valence, urgency, and motivational relevance, that shaped action long before deliberation was possible \cite{barrett2006emotions}. In this view, emotion is not an add-on to cognition but a foundational structure from which cognition evolved \cite{seth2016active}. This foundational role of emotion suggests that cognition did not evolve independently, but rather emerged through recursive interactions with emotion, forming a dynamic inference cycle in which affective structure continuously guides and is refined by cognitive specificity \cite{seth2013interoceptive}.

We propose that the recursive interaction between emotion and cognition \cite{pessoa2008emotion} forms a dynamic inference cycle that underlies adaptive intelligence. Emotion supplies high-level structural priors \cite{adolphs2018neuroscience}, while cognition instantiates specific beliefs or action plans consistent with affective goals \cite{pessoa2013emotion}. This \emph{emotion–cognition cycle} enables an agent to maintain coherence between internal states and external conditions by dynamically aligning context and content. Critically, this architecture supports both forward inference (inferring emotional meaning from external cues) and \emph{inverted inference} (generating plausible contexts from desired internal states) \cite{friston2010free, friston2017active}. The latter allows the system to simulate goal-consistent outcomes by modulating affective precision, thereby breaking the curse of dimensionality through constraint-based compression \cite{seth2013interoceptive, clark2013whatever}.

Building on this formulation, we present a formal account of \emph{consciousness as cycle-consistent affective inference}, wherein subjective experience emerges from recursive alignment between emotionally grounded goals and contextually updated beliefs \cite{tononi2004information}. When these cycles are temporally extended and information-theoretically coherent, they instantiate a latent variable, an \emph{affective self-model}, that binds diverse episodes into a unified structure \cite{seth2018being}. This mechanism explains how the system not only simulates the external world but also infers its own role and identity within it, achieving first-person experience through affective anchoring and recursive self-modeling \cite{dehaene2006conscious, carruthers2003phenomenal}.

We further extend this model through the lens of \emph{exchangeability}, as formalized in De Finetti’s representation theorem. By treating emotion–cognition cycles as conditionally exchangeable observations, we derive a stable latent self-model that integrates temporally diverse internal states into a coherent structure. This framework, termed the \emph{Exchangeable Integration Theory of Consciousness (EITC)}, bridges probabilistic inference and dynamical systems, unifying information integration, self-modeling, and consciousness. The EITC offers a unified framework in which consciousness emerges from the recursive alignment of emotion and cognition across time. By modeling emotion–cognition cycles as conditionally exchangeable episodes, EITC formalizes the construction of a latent affective self-model that binds diverse experiences into a coherent structure. 

Drawing inspiration from De Finetti’s representation theorem \cite{deFinetti1937}, EITC explains how the brain integrates temporally distributed, affectively weighted inferences under uncertainty. This integration not only preserves differentiation across episodes but also supports global coherence and recursive self-modeling \cite{griffin1994models}. We make a connection with the Bayesian theory of consciousness \cite{seth2022theories} by incorporating \emph{Rao-Blackwellization} as a mechanism for structure-preserving inference. In this view, consciousness arises from recursively refining internal models through the marginalization of latent self-structure while maintaining adaptive updates in response to contextual variation \cite{bellingrath2024emergence}. Through the lens of Rao-Blackwellized exchangeable integration, consciousness is not a static feature but a dynamical process of structure-preserving inference, one that integrates information, as in Integrated Information Theory (IIT) \cite{tononi2004information} and maintains accessibility, as in Global Workspace Theory (GWT) \cite{baars2005global}. 
The contributions of this paper are summarized below.

\begin{enumerate}
    
    \item \textbf{Emotion-cognition cycle supports bidirectional dynamics.} Emotion precedes cognition, which is consistent with the principle of structure-before-specificity. During the formation of the emotion-cognition cycle, emotion serves as an inferential structure, and cognition provides content specificity. The structural role of emotion explains why affect is indispensable to rationality and goal-directed action.
    
    \item \textbf{Inverted inference breaks the curse of dimensionality.} Instead of simulating forward from the present, inverted inference begins with the goal and infers backward the sequence of states and actions most likely to achieve it. Inverted inference turns a computational liability into an advantage: high-dimensional latent spaces offer the expressive capacity to encode rich goal hierarchies.

    \item \textbf{Content binding leads to the emergence of self.} Content binding in inverted inference naturally leads to the emergence of a latent variable that captures the agent’s internal structure. In De Finetti’s terms, this latent variable becomes the self-model: a hidden, recursively inferred entity ensuring subjective coherence across episodes of cognition and emotion.

    \item \textbf{Exchangeable integration unifies IIT and GWT of consciousness.} The EITC framework proposes that consciousness arises from conditionally exchangeable emotion–cognition cycles that construct a coherent latent self-model over time. EITC models consciousness as a recursive, affectively grounded inference process, integrating differentiated episodes into a temporally coherent structure through \emph{Rao-Blackwellized inference}, which marginalizes over latent self-structure to support both global integration and local contextual adaptation. 

\end{enumerate}


\section{Emotion Precedes Cognition}
\label{sec:2}

Emotion is among the earliest-evolved systems for adaptive behavior \cite{ledoux2012rethinking}. Long before the emergence of cognition, organisms developed affective circuits to regulate interaction with uncertain and dynamic environments. These circuits functioned to evaluate biological salience, prioritize sensory input, and guide action selection in real time. Emotion evolved as a survival mechanism that enabled efficient and flexible responses to environmental challenges without explicit reasoning \cite{panksepp1998affective}.
Comparative neuroscience reveals that core emotional systems, including fear, reward seeking, aggression, care, and play, are conserved across a wide range of vertebrate species \cite{panksepp1998affective}. These systems are implemented in subcortical structures such as the amygdala, hypothalamus, and periaqueductal gray, and they evolved well before the neocortex \cite{maclean1990triune}. Therefore, emotion provides a foundational control architecture that predates and scaffolds higher-order cognition.

From the perspective of statistical inference, emotion functions as the brain's original mechanism for inference under uncertainty. It compresses high-dimensional sensory and interoceptive data into low-dimensional, action-relevant states that encode valence, urgency, and motivational significance \cite{oosterhof2008functional}. These emotional states serve as \emph{structural priors} that constrain perceptual interpretation and behavioral responses.
As cortical structures expanded over evolutionary time, particularly in mammals and primates, cognition did not replace emotion; rather, it emerged through recursive integration with it \cite{pessoa2017network}. The prefrontal cortex co-evolved with limbic structures to support the cognitive regulation of emotion and, reciprocally, the emotional modulation of cognition \cite{barrett2006emotions}. This recursive coupling laid the groundwork for affective self-modeling, goal simulation, and ultimately, conscious awareness \cite{friston2020sentience}. 

A foundational principle in hierarchical inference is that \emph{structure precedes specificity} \cite{clark2013whatever}: global constraints guide the selective activation of local detail. This idea is reflected across multiple domains, from predictive coding in neuroscience \cite{friston2010free} to hierarchical planning in AI \cite{sacerdoti1974planning}. Recently, this principle has been extended into the context-content uncertainty principle (CCUP) \cite{li2025CCUP}, which frames cognition as a cyclic process of uncertainty minimization. Let \( \Psi \) denote context variables (e.g., ambiguous sensory input, environmental cues), and \( \Phi \) denote content variables (e.g., latent causes, goal states, task-relevant representations). CCUP asserts a fundamental trade-off in inference: the system seeks to minimize both 1) \textbf{specification/disambiguation}: conditional entropies \( H(\Phi | \Psi) \) (inference/encoding); and 2) \textbf{generalization/reconstruction}: \( H(\Psi | \Phi) \)  (generation/decoding). The main result of CCUP is stated as follows:

\begin{lemma}[CCUP Lower Bound]
Let \( \Psi \) and \( \Phi \) be random variables representing context and content. Then, we have
$H(\Phi | \Psi) + H(\Psi | \Phi) \geq H(\Psi, \Phi) - I(\Psi; \Phi)$
with equality if and only if the joint distribution \( p(\Psi, \Phi) \) is cycle-consistent and bidirectionally minimal.
\end{lemma}


The CCUP lower bound reveals a broken symmetry in cognition \cite{li2025Broken}: context variables \( \Psi \) are high-entropy and ambiguous, while content variables \( \Phi \) are lower-entropy and selectively encoded. This asymmetry implies that intelligent systems cannot infer specific content directly from raw context without incurring high uncertainty or representational interference. Instead, the system must first compress across multiple contexts to construct a generalizable structure that captures latent invariants \cite{bengio2013representation}. Only then can it safely bind specific experiences to that structure. This two-stage process, first minimizing \( H(\Psi|\Phi) \) to extract structure, then minimizing \( H(\Phi| \Psi) \) for context-sensitive binding, forms a robust strategy for reducing uncertainty across asymmetric inference cycles. It reflects a core principle of learning and memory in intelligent systems: \textbf{structure precedes specificity} (a.k.a. generalization-before-specification) \cite{clark2013whatever}.

In this work, we apply this principle to the interaction between emotion and cognition. We argue that \textbf{emotion serves as inferential structure}, while \textbf{cognition provides content specificity}. Emotion defines a low-dimensional, high-entropy prior over possible interpretations of internal and external context, dynamically shaping which cognitive contents are likely to be activated or suppressed. Conversely, cognition refines and specifies inference by selecting detailed hypotheses, appraisals, or actions within the affectively modulated landscape \cite{joffily2013emotional}. 
An emotion-cognition cycle reflects an evolutionary continuity in which affective structure is preserved and expanded through cortical recursion.
Formally, we start with the construction of the emotion-cognition cycle under the framework of CCUP \cite{li2025CCUP}.

\begin{definition}[Emotion-Cognition Cycle]
Let \( \Psi \) denote a high-entropy contextual prior associated with emotionally valenced states and \( \Phi \) denote low-entropy content representations instantiated by cognitive inference. The \emph{emotion–cognition cycle} is defined as a recursive inference process governed by the following dynamics:
1) \textbf{affective structuring:} Emotion imposes a global constraint on inference by shaping the prior distribution over latent variables: $p(\Phi | \Psi_{\text{emotion}})$, where \( \Psi_{\text{emotion}} \sim P(\Psi) \) encodes affectively biased contextual information;
2) \textbf{cognitive specialization:} Cognition performs a context-dependent posterior inference: $q(\Phi) = \arg\min_{q} \mathbb{E}_{q(\Phi)}\left[ \mathcal{L}(\Phi, \Psi) \right] + \beta D_{\mathrm{KL}}(q(\Phi) \Vert p(\Phi |\Psi))$, refining the content representation based on affectively modulated priors;
3) \textbf{feedback modulation:} The cognitive outcome \( \Phi^\ast \) updates the emotional context via an appraisal mechanism: $\Psi_{t+1} = f_{\text{affect}}(\Phi^\ast, \Psi_t)$, forming a recurrent loop that evolves over time.
\end{definition}

\noindent \textbf{Remark:} The cycle is said to be \emph{aligned} if it achieves joint uncertainty minimization:
$H(\Phi | \Psi) + H(\Psi | \Phi) \leq \varepsilon$,
for some bounded error tolerance \( \varepsilon \geq 0 \), and if the mutual information satisfies:
$I(\Psi; \Phi) \approx H(\Psi) \approx H(\Phi)$.
In other words, emotion functions as a \emph{global constraint generator}, modulating precision and determining which parts of the hypothesis space should be explored or ignored. Cognition functions as a \emph{local selector}, instantiating specific representations that minimize uncertainty under the given structure \cite{seth2016active}. This leads to the following lemma:

\begin{lemma}[Emotion as Structure, Cognition as Specificity]
Let \( \Psi_{\text{emotion}} \) denote an affective state encoding high-level contextual structure, and let \( \Phi_{\text{cognition}} \) denote specific cognitive content or hypotheses. Under the structure-before-specificity principle, the inference process is governed by:
$\Phi_{\text{cognition}} \sim p(\Phi | \Psi_{\text{emotion}})$
such that emotional states constrain the hypothesis space of cognition by modulating the conditional entropy:
$\min_{\Phi} H(\Phi | \Psi_{\text{emotion}})$,
where emotion sets a structural prior that reduces the search space of cognitive inference, while cognition instantiates specific beliefs or actions aligned with the affective context. 
\end{lemma}

\noindent \textbf{Remark:} This formulation supports the view that emotion and cognition are not independent modules, but coupled components in a recursive inference loop. The structural role of emotion explains why affect is indispensable to rationality and goal-directed action: it shapes what the system treats as relevant, plausible, or worth pursuing \cite{pezzulo2015active}. Next, we first develop a computational foundation for structural modeling of emotion and then extend it to include cognition as specificity. 

\section{Emotion-Cognition Cycle: Bidirectional Dynamics Under Uncertainty}
\label{sec:3}

Emotion and cognition are not separate systems but deeply intertwined processes that co-regulate adaptive behavior \cite{ochsner2007emerging}. Rather than treating emotion as secondary, contemporary neuroscience recognizes that emotion and cognition form a \emph{bidirectional inference cycle} under uncertainty, each modulating the other across attention, memory, and decision-making.

\subsection{Emotion as Structure: Toward an Inference Scaffold}
\label{sec:3.1}

Emotion, traditionally viewed as a reactive output of cognitive processes, is increasingly recognized as a structural component of inference and decision-making \cite{barrett2017emotions}.  In this framework, emotion serves as a global constraint generator, dynamically sculpting the space of plausible cognitive or behavioral responses, while cognition acts as a local selector, instantiating specific representations and actions within the affectively defined structure \cite{pezzulo2015active}. Computational models grounded in active inference and variational free energy minimization formalize this global-local asymmetry: emotional states are treated as latent modulators that influence the geometry of generative models and the curvature of the belief manifold \cite{smith2019simulating}. This structural perspective on emotion not only accounts for its early evolutionary origins but also provides a normative account of how affect scaffolds perception, memory, and action under uncertainty \cite{pezzulo2015active2}.

\begin{theorem}[Emotion as a Structural Prior]
Let \( \mathcal{H} \) denote the hypothesis space over latent causes \( h \), and let \( e \in \mathcal{E} \) be a low-dimensional affective state. Suppose inference proceeds by minimizing variational free energy:
$\mathcal{F}(q(h)) = \mathbb{E}_{q(h)}[-\log p(x | h, e)] + D_{\mathrm{KL}}(q(h) \,\Vert\, p(h | e))$.
Then we have:
\begin{enumerate}
    \item Emotion \( e \) defines a structural prior \( p(h | e) \) that constrains the hypothesis space \( \mathcal{H}_e \subseteq \mathcal{H} \), biasing inference toward affectively relevant regions.
    \item The affect-modulated posterior \( q(h) \) minimizes uncertainty locally under this structured prior:
    $q^*(h) = \arg\min_{q} \mathcal{F}(q(h); e)$.    
    \item Precision modulation \( \pi(e) \) further controls the weighting of prediction errors, shaping attentional gain and learning rate:
    $\mathcal{F}_\pi = \pi(e) \cdot \lVert x - \hat{x} \rVert^2 + D_{\mathrm{KL}}(q(h) \,\Vert\, p(h | e))$.   
\end{enumerate}
\label{thm:emotion_SP}
\end{theorem}

\noindent \textbf{Remark:} The proof of the above theorem is referred to the Appendix.
While emotion constrains inference by acting as a structural prior over latent hypotheses, it also serves a complementary computational role: projecting high-dimensional sensory and interoceptive data into low-dimensional affective states that guide hierarchical planning and goal decomposition.
Rather than encoding specific content, emotional states function as low-dimensional, high-entropy priors that shape the architecture of cognition itself \cite{oosterhof2008functional}. These priors modulate precision, guide attentional allocation, and constrain the hypothesis space by biasing inference toward affectively salient interpretations \cite{barrett2006emotions}. Formally, we have

\begin{corollary}[Emotion as a Low-Dimensional Projection for Hierarchical Planning]
Let \( s_t = (s^{\mathrm{ext}}_t, s^{\mathrm{int}}_t) \in \mathbb{R}^n \) denote the concatenated exteroceptive and interoceptive input at time \( t \), and let \( e_t \in \mathbb{R}^k \) be the corresponding affective state with \( k \ll n \). Suppose \( e_t = f_\theta(s_t) \) is a learned low-dimensional encoding that preserves behaviorally relevant structure.
Then:
\begin{enumerate}
    \item Emotion acts as a nonlinear dimensionality reduction function:
    $f_\theta: \mathbb{R}^n \rightarrow \mathbb{R}^k, ~ \text{where } e_t = f_\theta(s_t)$ 
    such that \( e_t \) encodes latent features of valence, urgency, and motivational significance relevant to action selection.

    \item The affective state \( e_t \) defines a structured prior over latent trajectories \( \{ h^{(l)}_t \} \), modulating the planning hierarchy as:
    $\forall l,~ p(h^{(l)}_t | h^{(l+1)}_t, e_t)$  

    \item The resulting planning objective becomes:
    $\mathcal{F}_{\mathrm{plan}} = \sum_{l=1}^{L} \sum_{t=1}^{T} \mathbb{E}_{q(h^{(l)}_t)} \left[ -\log p(x_t | h^{(l)}_t) \right] + D_{\mathrm{KL}}(q(h^{(l)}_t) \,\Vert\, p(h^{(l)}_t | e_t))$   
    
\end{enumerate}
\end{corollary}
\noindent \textbf{Remark:} Emotion serves not only as a structural prior but also as an information-theoretic bottleneck \cite{tishby1999information} that filters the high-dimensional flux of sensory and interoceptive inputs into low-dimensional latent states encoding valence and motivational salience \cite{tenenbaum2000global}. These compact affective representations act as task-relevant constraints, enabling tractable planning over abstract cognitive hierarchies \cite{niv2019learning}. By collapsing vast perceptual input into goal-relevant affective coordinates, emotion allows the agent to allocate cognitive resources efficiently and initiate appropriate behavioral policies. However, this bottom-up affective guidance must be balanced by top-down modulation to prevent overreactivity or maladaptive generalization \cite{keltner1999functional}. Cognition plays a critical counter-role: through recursive, goal-conditioned inference, cognitive systems can regulate, reinterpret, or suppress emotional responses \cite{tyng2017influences}. Next, we formalize this dynamic as \emph{inverted inference}, whereby cognition updates or reshapes the affective prior in light of contextual goals and learning history.

\subsection{Cognition Regulates Emotion via Inverted Inference}
\label{sec:3.2}

The perception–action cycle, traditionally understood as a loop between sensory input and motor output \cite{fuster2004upper}, is reinterpreted under the framework of inverted inference as a generative simulation process. Rather than passively reacting to stimuli, the agent begins with internal goals or affectively weighted priors and infers the most plausible actions and percepts that would fulfill them \cite{friston2017active}. In this inverted direction, inference flows backward, from outcomes to causes, allowing the agent to simulate potential futures, evaluate counterfactual trajectories, and plan goal-consistent behavior \cite{kaplan2018planning}. This reframing transforms perception into a hypothesis-testing mechanism and action into an inference-consistent update to the world, breaking the curse of dimensionality by a backward decomposition of global objectives \cite{bellman1966dynamic}. By treating both perception and action as recursive inference steps under internal constraints, the asymmetric cycle becomes a structure-preserving loop that minimizes uncertainty through active sampling of the environment. Crucially, this cycle is modulated by emotion, shaping the priors and precision that determine which hypotheses are explored and which actions are executed \cite{okon2015neurobiology}.

\begin{lemma}[Perception-Action Cycle as Inverted Inference]
Let \( \Phi_t \) be the latent state inferred from current perception \( \Psi_t \), and let \( \Phi_{\text{goal}} \) be an internally generated target state. We can characterize the perception-action cycle by:
\begin{enumerate}
    \item The agent infers actions as latent causes of future perceptual states:
    $a_t^* = \arg\max_a p(\Phi_{\text{goal}} | \Phi_t, a)$.
    
    \item This reverses the causal loop by initiating control from internally seeded priors rather than external stimuli.

    \item The action policy minimizes expected conditional entropy:
    $a_t^* = \arg\min_a H(\Phi_{\text{goal}} | \Phi_t, a)$
\end{enumerate}
\end{lemma}

\noindent \textbf{Remark:}
The perception-action cycle constitutes an inverted inference process, wherein the agent selects actions that maximize the likelihood of achieving internally defined goals, rather than merely reacting to external inputs \cite{li2025Arrow}. Such goal-directed simulation offers an important insight into breaking the curse of dimensionality based on Bellman's principle of optimality \cite{bellman1966dynamic}. Formally, we have

\begin{proposition}[Inverted Inference Facilitates Goal-Directed Simulation]
Let \( \Phi_{\text{goal}} \) be a low-entropy goal state in the latent space, and let \( \Phi_{0:T} \) be a trajectory of latent states inferred by simulation. Then inverted inference enables efficient goal simulation by:
1) {Reversing the direction of inference:}
    $p(\Phi_{0:T} | \Phi_{\text{goal}}) \ll p(\Phi_{0:T} | \Phi_0)$;    
2) {Minimizing conditional entropy of the simulated path:}
    $H(\Phi_{0:T} | \Phi_{\text{goal}}) \ll H(\Phi_{0:T} | \Phi_0)$;
3) {Restricting inference to goal-relevant trajectories:}
    $\Phi_{0:T} \in \mathcal{M}_{\text{goal}}, ~ \dim(\mathcal{M}_{\text{goal}}) \ll \dim(\Phi)^T$. 
\end{proposition}

\noindent \textbf{Remark:}
Note that inverted inference transforms the task of goal simulation from an intractable forward search problem into a structured, goal-conditioned inference process over a compressed subset of the latent state space (goal-directed manifold) \cite{pezzulo2014internally}. Such manifold constraint effectively reduces the search space by turning a computational liability into an advantage: high-dimensional latent spaces offer the expressive capacity
to encode rich goal hierarchies \cite{ha2018world}. This compression of the search space reveals a deeper insight: the expressive power of latent manifolds is co-regulated by emotion and cognition, whose interaction forms a co-adaptive engine of intelligence and consciousness \cite{pessoa2008emotion}.
Rather than opposing faculties, emotion and cognition form a co-adaptive system whose interaction is central to intelligence and consciousness. Therefore, we have

\begin{proposition}[Emotion as Contextual Prior in Selective Inference]
Let \( \Psi_{\text{emotion}} \) denote an emotional state encoding high-entropy, low-dimensional contextual information, and \( \Phi_{\text{content}} \) denote cognitive content representing low-entropy, high-dimensional mental representations. Emotional states function as contextual priors that constrain the inference over cognitive content under uncertainty by
$\min_{\Phi} H(\Phi | \Psi_{\text{emotion}})$,
leading to selective activation of content \( \Phi \) that is congruent with the emotional context \( \Psi_{\text{emotion}} \), effectively reducing the inference space and facilitating rapid, goal-aligned cognition.
\end{proposition}

\noindent \textbf{Remark:} The proof of the above theorem is referred to the Appendix.
Under the predictive coding framework \cite{keller2018predictive}, let \( \hat{\Phi} \) denote top-down predictions and \( \epsilon = \Psi - \hat{\Psi} \) denote bottom-up prediction errors. Emotional states modulate the precision (inverse variance) \( \pi(\Psi_{\text{emotion}}) \) assigned to contextual prediction errors. This modulation alters the update rule for beliefs about \( \Phi \) as follows:
$\Delta \hat{\Phi} \propto \pi(\Psi_{\text{emotion}}) \cdot \epsilon$.
Therefore, emotional priors not only bias content selection but also regulate the gain on incoming signals, influencing how strongly prediction errors update cognitive beliefs \cite{seth2013interoceptive}.
High-arousal emotions (e.g., fear, anger) increase precision \( \pi \), enhancing sensitivity to prediction errors and accelerating inference updates. Low-arousal or neutral states reduce precision, promoting exploratory or integrative cognition \cite{barrett2017theory}. This precision weighting enables adaptive control of the balance between stability (prior-driven) and flexibility (error-driven) cognition \cite{mermillod2013stability}. Formally, we have (its proof can be found in the Appendix)

\begin{theorem}[Content Binding via the Emotion–Cognition Cycle]
Let \( e \in \mathcal{E} \) be an emotional state that regulates both inference and precision weighting.
The emotion–cognition cycle operates through two coupled inferential processes:
1) \textbf{forward (reactive) inference:} emotional context shapes cognitive content via $\Phi_t \sim p(\Phi | \Psi_t, e_t)$, filtering and binding perceptual and interoceptive features into affectively relevant content; 2) \textbf{inverted (goal-directed) inference:} desired emotional or cognitive states \( \Phi^* \) induce simulations over plausible contexts: $\Psi_{t+1} \sim p(\Psi | \Phi^*, e_t)$,   
enabling proactive planning and counterfactual inference.
The recursive interaction between these processes forms a temporally extended inference cycle:
$\Phi_t \xleftrightarrow{e_t} \Psi_{t+1} \xleftrightarrow{e_{t+1}} \Phi_{t+1}$,
in which each content binding \( \Phi_{t+1} \) is constrained by prior emotional states and contextual predictions. This cycle minimizes joint uncertainty \( H(\Phi, \Psi) \), stabilizes feature integration across time, and constructs coherent internal representations grounded in affectively weighted relevance. 
\label{thm:content_binding}
\end{theorem}
\noindent \textbf{Remark:} Emotion acts as a dynamic binding prior, ensuring that cognitive content remains structured, goal-consistent, and context-sensitive.
Acting as a temporally stable, low-dimensional prior, emotion dynamically sculpts the landscape of relevance, shaping which features of the internal and external environment are selected for processing, remembered, or acted upon \cite{mather2011arousal}. Within this framework, \emph{content binding}, the process by which perceptual, interoceptive, and conceptual features are integrated into unified representational structures, is stabilized by affectively salient contents \cite{treisman1996binding}.

\subsection{Self-Awareness from Emotion-Cognition Cycle}
\label{sec:3.3}

The perception–action cycle, when reformulated as inverted inference, reveals how cognition emerges through recursive simulation of goal-consistent trajectories \cite{li2025Arrow}. Yet, this mechanism alone does not account for the subjective continuity or identity that underlies self-awareness. To address this, we must consider how emotionally valenced priors modulate and stabilize the inference process over time \cite{bellingrath2024emergence}. It is through the recursive interaction of emotion and cognition that contents are not only selected but also bound into coherent episodes that reflect a stable point of view. This transition, from goal-conditioned inference to identity-conditioned coherence, marks a crucial shift: the emergence of a latent self-model \cite{griffin1994models}. In what follows, we explore how content binding across emotion–cognition cycles gives rise to self-awareness, framed as the construction of a temporally persistent generative prior over conscious experience \cite{seth2018being}.

Let \( e_t \in \mathcal{E} \) denote the emotional context at time \( t \), and let \( \Phi_t \in \mathcal{F} \) represent the set of bound content features selected under this context. Over time, the system accumulates a sequence \( \{\Phi_t\}_{t=1}^T \) of emotionally modulated representations, which serve as evidence for a latent variable \( \theta \in \Theta \) corresponding to the inferred self-model. The system performs inference over \( \theta \) using a temporally recursive Bayesian update:
\begin{equation}
q(\theta) \propto p(\theta) \prod_{t=1}^{T} p(\Phi_t | \theta, e_t),
\label{eq:1}
\end{equation}
where \( p(\Phi_t | \theta, e_t) \) is the likelihood of observing a particular content binding given the current emotional state and hypothesized self-model. Emotion not only modulates the binding process (i.e., which contents are integrated \cite{edelman2008universe}), but also influences the posterior distribution over \( \theta \), effectively shaping the emergent identity of the agent.
This recursive construction gives rise to a \textbf{latent self-model} \cite{griffin1994models}: a structured, probabilistic representation of the agent’s identity, inferred from emotionally constrained episodes of content binding. It reflects both the continuity of subjective experience and the adaptive plasticity required to align future inference with internal goals. Thus, emotion and cognition jointly construct a temporally extended model of the self, grounded in affectively weighted interaction with the world \cite{seth2013interoceptive}. Formally, we conclude this section with the following theorem.

\begin{theorem}[Latent Self-Model from Content Binding]
Let \( \{ (\Psi_t, \Phi_t) \}_{t=1}^T \) denote a temporally ordered sequence of context–content pairs generated through an inverted inference process, where each \( \Phi_t \sim q(\Phi | \Psi_t) \) is affectively constrained and content-bound. Suppose the system exhibits cycle-consistent alignment:
$D_{\mathrm{KL}}(P(\Psi_t) \, \Vert \, p(\Psi_t | \Phi_t)) \leq \varepsilon ~ \text{and} ~ I(\Psi_t; \Phi_t) \approx H(\Psi_t)$
for all \( t \in \{1, \dots, T\} \) and some \( \varepsilon \geq 0 \). Then, under the assumption of informational coherence across episodes, there exists a latent variable \( \theta \in \Theta \) such that:
\begin{equation}
   P(\Phi_1, \dots, \Phi_T) = \int_\Theta \prod_{t=1}^T P(\Phi_t | \theta) \, dP(\theta), 
\end{equation}
where \( \theta \) functions as a latent self-model that unifies content binding under a common generative structure.
\label{thm:latent_self_model}
\end{theorem}
\noindent \textbf{Remark:} The emergence of a latent self-model from temporally structured, emotionally guided content binding offers a powerful bridge to a new computational theory of consciousness \cite{dehaene2014toward}. As formalized in the preceding theorem, the affectively constrained cycle between context and content produces a sequence of cognitive-emotional states whose statistical regularities are captured by a latent variable $\theta$, interpretable as the self. This structure mirrors De Finetti’s theorem \cite{deFinetti1937}, wherein exchangeable sequences of observations imply the existence of an underlying generative cause. 

Under the CCUP framework, consciousness is not a static entity but a dynamically inferred construct, a temporally extended, recursively updated self-model anchored by emotion and refined through cognition. The emotion–cognition cycle, by regulating both inference and content binding, serves as the computational substrate for constructing and maintaining subjective continuity \cite{seth2013interoceptive}. This view reframes consciousness as the informational footprint of recursive, affectively modulated inference processes that unify diverse episodes of perception and action into a coherent model of the self-in-context.

\section{Emotion-Cognition Cycle and the Hard Problem of Consciousness}
\label{sec:4}

The ``hard problem of consciousness'' asks why and how physical or computational processes give rise to subjective experience, the \emph{qualia} or felt sense of being \cite{buzsaki2006rhythms}. While functionalist and neuroscientific accounts have addressed the mechanisms of perception, attention, and awareness, they often fall short of explaining \emph{why} certain brain states are accompanied by experience \cite{edelman2008universe}. We propose that the \emph{emotion-cognition cycle}, as formalized under CCUP \cite{li2025CCUP}, offers a computational pathway that narrows this explanatory gap. In this section, we will present three complementary perspectives toward consciousness (i.e., phenomenological consciousness \cite{carruthers2003phenomenal}, access consciousness \cite{block1995confusion}, and self-consciousness \cite{bermudez2000paradox}), in the order of increasing sophistication and technical depth, to illustrate that {\bf consciousness is a temporally extended, recursive inference process that minimizes uncertainty through affectively modulated alignment between context and content}.

\subsection{Phenomenological Consciousness as Cycle-Consistent Affective Inference} 
\label{sec:4.1}

Phenomenological consciousness refers to the qualitative, first-person character of experience—commonly described as {\em ``what it is like''} to be in a particular mental state. This concept has roots in both philosophy of mind and phenomenology, where it is distinguished from unconscious or merely functional processes. As introduced by \cite{nagel1980like}, any conscious system must have some internal perspective or point of view. In contemporary cognitive science, phenomenological consciousness is increasingly studied through its neurocomputational underpinnings, particularly in frameworks like predictive processing \cite{keller2018predictive} and active inference \cite{friston2020sentience}. 

In this work, conscious experience is modeled as the informational footprint of recursive perceptual inference that integrates interoceptive and exteroceptive signals under uncertainty \cite{seth2013interoceptive}. Emotion plays a central role in this process by shaping the precision weighting and salience of experience, effectively grounding phenomenological content in affectively modulated dynamics \cite{craig2004human}. Rather than being epiphenomenal, phenomenological consciousness may thus emerge from an integrated loop where emotion constrains inference and inference refines emotional structure over time. Under the CCUP framework, we propose the following definition.

\begin{definition}[Phenomenological Consciousness]
Let \( \{ (\Psi_t, \Phi_t) \}_{t=1}^T \) be a temporally ordered sequence of context–content pairs generated via an emotion–cognition cycle, where \( \Phi_t \sim q(\Phi | \Psi_t, e_t) \) and \( e_t \in \mathcal{E} \) denotes an affective state modulating inference. Suppose this sequence is cycle-consistent and admits a latent variable \( \theta \in \Theta \) such that:
$P(\Phi_1, \dots, \Phi_T) = \int_\Theta \prod_{t=1}^T P(\Phi_t | \theta) \, dP(\theta)$,
where \( \theta \) functions as a temporally integrated self-model.
We define \emph{Phenomenological Consciousness} as the emergent property of this recursive process, characterized by: 1) \textbf{Affective Grounding}: Each \( \Phi_t \) is shaped by emotionally weighted relevance filtering via \( e_t \), determining salience and precision; 2) \textbf{Content Binding}: Contextually consistent integration of multimodal features into structured representations \( \Phi_t \); 3) \textbf{Temporal Coherence}: Inference of a shared latent self-model \( \theta \) across episodes ensures subjective continuity.
\end{definition}

Phenomenological consciousness is the informational footprint of a system that recursively infers and binds affectively modulated experiences into a coherent, temporally extended self-representation. Using the emotion-cognition cycle as a scaffold, we decompose a computational modeling of phenomenological consciousness into the following three building blocks: dynamical inference, emotion regulation, and intentionality generation.

\noindent\paragraph{Dynamic Inference as Conscious Substrate}
Let \( \Phi \) represent internal content (e.g., goals, beliefs, desires), and \( \Psi \) represent contextual or sensory variables. The emotion-cognition cycle describes an iterative loop:
$\Phi_t \leftrightarrow \Psi_{t+1} \leftrightarrow \Phi_{t+1}$
This cycle, actively aligning internal cognitive-emotional content with perceived or imagined context,
is seeded by goal-conditioned priors \( \Phi^* \) and modulated by emotion, which adjusts the precision weighting of context-dependent inference.
Subjective experience, the core of consciousness, is proposed to emerge from this cycle when it exhibits sufficient temporal depth, self-referential structure, and affective modulation.

\noindent\paragraph{Emotion as Phenomenological Scaffold}

Emotion plays a dual role: 1) As a \emph{modulator of precision}, emotion influences the salience and reliability of contextual inputs.
 2) As a \emph{self-modeling signal}, it provides recursive access to how the system values its own states, forming the basis of affective awareness.
Recursive emotional appraisal (e.g., ``how do I feel about how I feel'') leads to meta-cognitive loops, core constituents of phenomenological depth.

\noindent\paragraph{Counterfactual Depth and the Feeling of Intentionality}
When the system performs \emph{inverted inference}, simulating contexts that would satisfy internally specified goals \( \Phi^* \): $\Psi \sim p(\Psi | \Phi^*)$, it generates counterfactual futures conditioned on affective priors. This gives rise to the felt sense of \emph{intention} or \emph{agency}, the experience of \emph{desire}, \emph{imagination}, or \emph{anticipation}, and
a self-referential inference process that is temporally extended.

Together, these three elements constitute the building blocks of conscious experience. 
Rather than eliminating the ``hard problem'' this account \emph{reframes} it: subjective experience becomes the \emph{informational footprint} of a system that is recursively bootstrapping coherence between content and context across affectively weighted inferential cycles \cite{melloni2021making}. Formally, we have

\begin{proposition}[Cycle-Consistent Inference and Phenomenological Consciousness]
Let \( \Phi \) be the internal content space, \( \Psi \) the context space, and \( \Phi^* \subset \Phi \) represent affectively valenced goal states. Under CCUP, a system exhibits phenomenological consciousness when it supports:
1) Forward inference: \( \Phi \sim p(\Phi | \Psi) \); 2) Inverted inference: \( \Psi \sim p(\Psi | \Phi^*) \); 3) Recursive alignment: \( \Phi_t \leftrightarrow \Psi_{t+1} \leftrightarrow \Phi_{t+1} \); and 4) Affective modulation of precision: \( \pi(\Psi) = f(\Phi^*) \), where \( \pi \) denotes precision weighting.
The joint entropy \( H(\Phi, \Psi) \) is minimized over time via dynamic cycle consistency, and the system exhibits temporally integrated, affectively grounded self-modeling. This structure constitutes the computational substrate for phenomenological consciousness.
\end{proposition}

\noindent\textbf{Remark:}
This framework suggests that consciousness is not a static state or binary property, but a \emph{process} emerging from recursive, emotion-modulated inference under uncertainty \cite{friston2017active}. Emotion provides the phenomenological \emph{texture}; recursive alignment provides the \emph{continuity}; and goal simulation under CCUP provides the \emph{intentional structure} of experience \cite{barrett2017theory}.

\subsection{Access Consciousness as Temporally-Extended Affect-Modulated Inference}
\label{sec:4.2}

Access consciousness refers to mental content that is not only experienced but also globally available for reasoning, verbal report, and behavioral control \cite{block1995confusion}. A classic case highlighting its distinction from phenomenal consciousness is \textbf{blindsight}, where individuals with damage to the primary visual cortex can respond to visual stimuli without reporting any conscious visual experience. Although these patients lack phenomenal awareness, their ability to make accurate decisions about unseen stimuli suggests that visual information is still accessible to cognitive systems. This dissociation indicates that conscious access involves a distinct mechanism, often modeled as a global workspace \cite{baars2005global}, in which selected information is broadcast across multiple functional modules, such as memory, attention, and motor planning. 

Theories of predictive processing and active inference interpret access consciousness as the result of successful model selection: emotionally relevant and inferentially robust content is promoted to global availability. Access consciousness emerges when information becomes structured, precision-weighted, and sufficiently context-sensitive to coordinate adaptive action and self-directed behavior \cite{seth2022theories}. Therefore,
access consciousness denotes the set of emotionally salient representations that are globally available for inference, memory, action, and decision-making. Such observation leads to the following definition.

\begin{definition}[Access Consciousness]
Let \( \Phi_t \) denote a content representation inferred at time \( t \), and \( \mathcal{G}_t \subseteq \mathcal{M} \) represent the set of globally accessible mental systems. A representation \( \Phi_t \) is said to \emph{access conscious} if it is 1) \textbf{emotionally relevant}: \( \Phi_t \sim q(\Phi | \Psi_t, e_t) \) for some emotionally weighted context \( \Psi_t \) and affective prior \( e_t \); 2) \textbf{globally broadcast}: \( \Phi_t \in \bigcap_{m \in \mathcal{G}_t} \text{Input}(m) \) (i.e., it is available as input to all systems in \( \mathcal{G}_t \)).
\end{definition}

While cycle-consistency captures the temporal coherence of consciousness, a deeper functional insight emerges when we recognize that the recursive interaction mediates this consistency between emotion and cognition, framing access consciousness not merely as sustained inference, but as affectively guided, cognitively refined simulation. Next, we propose a unifying theoretical framework in which access consciousness emerges from a temporally extended, recursive inference process modulated by the interaction between emotion and cognition. This view integrates concepts from active inference \cite{friston2017active}, optimal transport (OT) theory \cite{peyre2019computational}, and the information bottleneck method \cite{tishby1999information}. Consciousness is framed not as a static state, but as {\em a dynamic, cycle-consistent alignment between perceptual context and cognitive content over time for accessing internal goals}. We first construct temporally-extended inference by connecting with the theory of OT alignment \cite{peyre2019computational}.


\begin{definition}[Optimal Transport Alignment]
Let \( \pi(\Psi_{1:T} | \Phi^*) \) denote a goal-conditioned distribution over context trajectories induced by an affective anchor \( \Phi^* \), and let \( q(\Psi_{1:T}) \) denote the inferred distribution over contextual states. Define a ground cost function \( c : \Psi \times \Psi \rightarrow \mathbb{R}_{\geq 0} \), typically \( c(\psi, \psi') = \| \psi - \psi' \|^2 \). The Optimal Transport (OT) cost between the two distributions is given by:
\begin{equation}
\mathrm{OT}\left( \pi, q \right) := \inf_{\gamma \in \Gamma(\pi, q)} \int_{\Psi \times \Psi} c(\psi, \psi') \, d\gamma(\psi, \psi'),
\end{equation}
where \( \Gamma(\pi, q) \) is the set of all couplings (transport plans) \( \gamma \) over \( \Psi \times \Psi \) with marginals \( \pi \) and \( q \), respectively. This formulation captures the minimal cost required to transport the simulated goal-conditioned context trajectory into the inferred empirical context trajectory.
\end{definition}

\noindent\textbf{Temporally Extended Inference via Optimal Transport.}
Let \( \Phi^* \) represent a goal or affective anchor, and let \( \pi(\Psi_{1:T} | \Phi^*) \) denote a simulated trajectory of contexts consistent with that goal, a product of \emph{inverted inference}. Let \( q(\Phi_{1:T} | \Psi_{1:T}) \) represent the actual inferred trajectory of content under incoming context, a result of \emph{forward inference}.
We define the conscious inference process as a recursive optimization that minimizes the sum of context-content uncertainty and transport mismatch between simulated and actual trajectories:
$\min_{q} \left[ \sum_{t=1}^T \gamma_t H(\Phi_t | \Psi_t) + \lambda \, \mathrm{OT} \left( \pi(\Psi_{1:T} | \Phi^*), q(\Psi_{1:T}) \right) \right]$,
where \( \gamma_t \) encodes affective precision, modulating attention and inference strength based on emotion, \( H(\Phi_t | \Psi_t) \) is the conditional entropy of content given context at time \( t \), and \( \mathrm{OT}(\cdot,\cdot) \) is an OT cost (e.g., Wasserstein distance \cite{kantorovich1960mathematical}) aligning goal-simulated and observed trajectories.

Alternatively, we can frame this as a dynamic information bottleneck problem \cite{tishby1999information}. Consciousness arises from selecting representations that maximize relevance to immediate context while retaining coherence with affective structure:
$\mathcal{L}_t := I(\Psi_t; \Phi_t) - \beta I(\Phi_t; \Phi^*)$
This objective preserves content that is both contextually informative and emotionally aligned, enforcing a bounded, low-entropy conscious workspace.
To temporally extend the concept of phenomenological continuity, the system must recursively reconstruct internal states in a way that is consistent with prior structure and evolving context \cite{li2025Inverted}. This is formalized as a temporal consistency constraint:
$q(\Phi_{t+1}) \approx f(\Phi_t, \Psi_t), ~\text{and}~ \Psi_t \sim p(\Psi_t | \Phi^*)$,
which leads to a bidirectional inference loop that stabilizes over time and supports conscious awareness. Formally, we have

\begin{proposition}[Emotion-Cognition Cycle as the Basis for Access-Conscious Inference]
Let \( \Phi^* \) denote an affective anchor (emotion), \( \Phi_t \) the cognitive state at time \( t \), and \( \Psi_t \) the contextual input. Then consciousness arises from a temporally extended and affect-modulated inference process that recursively minimizes:
$\sum_{t=1}^T \gamma_t H(\Phi_t | \Psi_t) + \lambda \, \mathrm{OT}\left( \pi(\Psi_{1:T} | \Phi^*), q(\Psi_{1:T}) \right)$ 
subject to:
$q(\Phi_{t+1}) \approx f(\Phi_t, \Psi_t), \Psi_t \sim p(\Psi_t | \Phi^*)$
This process instantiates the emotion-cognition cycle as a recursive alignment between contextual input \( \Psi_t \), which updates cognitive states via forward inference, and cognitive content \( \Phi_t \), which recursively reconstructs goal-relevant trajectories, 
subject to an emotional structural constraint \( \Phi^* \), which modulates inference via affective precision \( \gamma_t \).
\end{proposition}
\noindent\textbf{Remark:} Access consciousness emerges as the fixed point of this bidirectional loop that stabilizes internal and external alignment over time. A cycle-consistent, affectively anchored process that maintains predictive coherence across self, world, and time \cite{seth2022theories}. While access consciousness ensures that emotionally salient and inferentially robust content becomes globally available for action and reasoning, self-consciousness arises when this content is recursively integrated into a coherent, temporally extended self-model that anchors perception, memory, and goal simulation. However, how to model the latent self-model has remained an open question absent from both models of phenomenal consciousness and access consciousness.

\subsection{Self-Consciousness as Exchangeability-Regularized Inverted Inference}
\label{sec:4.3}

While the emotion–cognition cycle accounts for the dynamic, recursive structuring of conscious experience, where emotionally weighted priors constrain cognitive specificity, it remains to explain how such a system achieves \textbf{coherence} across internally simulated possibilities. 
Modeling self-consciousness as a temporally extended inference process poses a central challenge: how can coherence be preserved across diverse, fluctuating experiences \cite{friston2018self}? Exchangeability offers a solution: if emotion–cognition cycles are conditionally exchangeable, then a stable latent structure (the self) can be inferred as the common cause underlying these affectively weighted interactions over time.

The notion of \textbf{exchangeability} offers a powerful extension of consciousness in time conceptually similar to the ergodic hypothesis \cite{niepert2014tractability}: if conscious inference is not about computing fixed truths, but about maintaining internal consistency across imagined, remembered, and perceived states, then the self must be treated as a latent variable governing an exchangeable sequence of experiences \cite{griffin1994models}. In other words, self-consciousness must reflect the recursive inference of a latent self-model that unifies emotional, cognitive, and contextual representations across time. Such observation leads to the following definition for self-consciousness.

\begin{definition}[Self-Consciousness]
Let \( \{ (\Psi_t, \Phi_t) \}_{t=1}^T \) be a temporally ordered sequence of context–content pairs aligned through an emotion–cognition cycle, and suppose there exists a latent variable \( \theta \in \Theta \) such that
$P(\Phi_1, \dots, \Phi_T) = \int_\Theta \prod_{t=1}^T P(\Phi_t | \theta) \, dP(\theta)$, where \( \theta \) encodes an internal self-model. A system is said to be \emph{self-conscious} if it maintains a temporally integrated latent model \( \theta \) inferred from experience and uses \( \theta \) to simulate counterfactuals involving itself. 
\end{definition}

Inspired by de Finetti’s representation theorem \cite{deFinetti1937}, this formulation enables the agent to infer a unified latent variable, interpretable as the self, that probabilistically organizes and binds otherwise disjoint experiences. The key motivation here is that self-consciousness does not require storing or processing each experience in full detail; instead, it requires recognizing that these experiences are structured by a stable generative source. Exchangeability thus provides the symmetry assumption necessary for recursive, self-refining inference \cite{griffin1994models}: by treating past and future observations as conditionally governed by the same latent self, the system can flexibly update its identity model while preserving continuity. Moreover, exchangeability allows a system to treat sequences of cognitive-emotional states as conditionally independent and identically distributed (i.i.d.) given a latent self-model. 

\noindent\paragraph{Latent Variable \texorpdfstring{$\theta$}{θ} as the Self}
De Finetti’s representation theorem establishes that any infinite sequence of exchangeable random variables can be represented as conditionally independent and identically distributed (i.i.d.) given a latent variable \( \theta \). Formally, for a sequence of random variables \( X_1, X_2, \ldots \), exchangeability implies the existence of a parameter \( \theta \) such that:
$P(X_1, \dots, X_n) = \int \prod_{i=1}^n P(X_i | \theta) \, dP(\theta)$.
This representation suggests that the identity of the sequence arises not from its observable values but from the invariant structure induced by \( \theta \). In the context of consciousness, we interpret \( \theta \) as the {\bf latent self-model} \cite{griffin1994models}, a hidden generative factor that ensures subjective coherence across perceptual, cognitive, emotional, and imagined experiences.

\noindent\paragraph{Interpretation in Affective Inference}
Within the emotion–cognition cycle, the system continually aligns content \( \Phi \) and context \( \Psi \) to minimize joint uncertainty. However, over time and across modalities, such alignment must be constrained by a persistent internal reference, an anchor that maintains structural coherence. We propose that this anchor corresponds to the latent variable \( \theta \), which encodes the agent’s stable self-structure. This latent self governs how experiences are interpreted, emotionally weighted, and stored/recalled \cite{seth2013interoceptive}.

\noindent\paragraph{Exchangeability as Subjective Coherence}
From a phenomenological standpoint, the subjective sense of continuity, the feeling that {\em “I am the same person across experiences”}, can be understood as a manifestation of exchangeability \cite{dainton2008phenomenal}. Just as exchangeable sequences admit a common generative model, subjective experiences are encoded, evaluated, and simulated under the assumption of a consistent self-model \( \theta \).
Let \( \{ \xi_t \}_{t=1}^T \) denote the temporally ordered sequence of conscious episodes. If this sequence is modeled as exchangeable under a latent self-variable \( \theta \), then
$P(\xi_1, \dots, \xi_T) = \int \prod_{t=1}^T P(\xi_t | \theta) \, dP(\theta)$,
where \( \theta \) acts as a unifying source of prior expectations, identity continuity, and affective valuation. This formulation supports the view that consciousness is not a static output of neural computation but an emergent, structured process of inference, anchored by a generative latent self that maintains coherence across recursive cycles of emotion and cognition. The above line of reasoning can be summarized into the following result.

Probing into self-consciousness from the lens of exchangeability aligns with the intuition that the self is not an object directly perceived, but a structured prior inferred through consistency across internally and affectively bound experiences.
In this view, self-consciousness emerges not from fixed identity or content, but from a subjective expectation of coherence across time and modality, a coherence achieved through Bayesian symmetry \cite{fenigstein1975public}. This bridges local emotion-cognition cycles with a global, probabilistic principle of consistency, setting the stage for modeling consciousness as exchangeable, affectively anchored inference.

\begin{theorem}[Conscious Experience as Exchangeable Inference]
Let \( X_{1:T} = (X_1, X_2, \dots, X_T) \) represent a temporally ordered sequence of internal states or observations that are conditionally exchangeable under a generative model. Based on De Finetti's theorem, there exists a latent variable \( \Phi^* \), interpreted as an affectively grounded self-model, such that:
$P(X_{1:T}) = \int \prod_{t=1}^T P(X_t | \Phi^*) \, dP(\Phi^*)$
This implies that:
\begin{enumerate}
    \item Conscious experience arises as the subjective inference over a latent self-structuring variable \( \Phi^* \) that renders internally sensed events conditionally i.i.d.;
    \item Emotion provides the structural prior \( P(\Phi^*) \), anchoring the generative model of self and affective relevance;
    \item Cognition instantiates the recursive update of posterior beliefs \( P(\Phi^* | X_{1:t}) \) as more observations accrue;    
\end{enumerate}
\label{thm:exchange_infer}
\end{theorem}
\noindent\textbf{Remark:}
The temporally extended coherence of conscious awareness corresponds to the exchangeability constraint, i.e., that subjective episodes are organized as if drawn from a common latent cause.
This formalizes self-consciousness as a Bayesian process of structure-preserving inference \cite{seth2022theories}, where emotion encodes the prior over latent causes and cognition recursively selects content consistent with that structure. Building on the insight that conscious experience emerges from exchangeable inference over a latent self-model, we now turn to how this statistical structure constrains simulation: \emph{exchangeability regularizes inverted inference} by enforcing coherence across imagined trajectories, ensuring that goal-directed simulations remain consistent with a unified generative self.

\begin{corollary}[Exchangeability Regularizes Inverted Inference]
Let \( \{X_t\}_{t=1}^T \) denote a sequence of internal states or observations, assumed to be exchangeable under a latent affective prior \( \Phi^* \), such that:
$P(X_{1:T}) = \int \prod_{t=1}^T P(X_t | \Phi^*) \, dP(\Phi^*)$.
The process of inverted inference, defined as simulating plausible context trajectories \( \Psi_{1:T} \) conditioned on a goal-anchored structure \( \Phi^* \),
$\pi(\Psi_{1:T} | \Phi^*) := \arg\min_{\Psi_{1:T}} \mathcal{C}(\Psi_{1:T}; \Phi^*)$
is regularized by exchangeability in the following sense:
\begin{enumerate}
    \item \textbf{Dimensionality reduction:} Exchangeability implies that inference can reuse the same latent cause \( \Phi^* \) across time, reducing the complexity of simulation;
    \item \textbf{Path coherence:} The simulated trajectory \( \pi(\Psi_{1:T} | \Phi^*) \) maintains consistency with both past and future states due to conditional i.i.d.-ness;
    \item \textbf{Cycle preservation:} The bidirectional inference loop
    $\pi(\Psi_{1:T} | \Phi^*)  \longleftrightarrow  q(\Phi^* | \Psi_{1:T})$
    remains cycle-consistent under the exchangeability prior;
    \item \textbf{Collapse avoidance:} The inference process avoids path integral collapse by preserving structural coherence across permutations of \( X_{1:T} \), thus preventing fragmentation or drift of the self-model.
\end{enumerate}
\end{corollary}
\noindent\textbf{Remark:}
Therefore, exchangeability functions as an implicit structural prior that anchors inverted inference, stabilizes the emotion–cognition cycle, and sustains the continuity of conscious simulation. By acting as a statistical symmetry constraint, exchangeability not only stabilizes inverted inference but also enables the construction of a coherent latent self-model: a common generative cause that organizes temporally dispersed experiences into a unified representational trajectory.

\begin{theorem}[Latent Self-Model from Exchangeable Experience]
Let \( \{X_t\}_{t=1}^T \) be a finite sequence of conscious states, observations, or internal variables. Suppose this sequence is exchangeable, i.e.,
$P(X_1, \dots, X_T) = P(X_{\pi(1)}, \dots, X_{\pi(T)})$
for any permutation \( \pi \in S_T \). Then, by De Finetti's representation theorem, there exists a latent variable \( \theta \in \Theta \) such that:
$P(X_{1:T}) = \int_\Theta \prod_{t=1}^T P(X_t | \theta) \, dP(\theta)$
If the agent maintains an internal self-model \( \Phi^* \in \Theta \) via Bayesian inference, then:
\begin{enumerate}
    \item The latent variable \( \theta \) corresponds to the agent’s \emph{affective self-model} \( \Phi^* \), providing a stable, time-invariant structure underlying subjective experience.
    \item The posterior update
    $P(\Phi^* | X_{1:T}) \propto P(\Phi^*) \prod_{t=1}^T P(X_t | \Phi^*)$
    defines recursive self-modeling consistent with predictive coding and active inference.
    \item Inverted inference is enabled by anchoring counterfactual simulations on \( \Phi^* \), i.e.,
    $\pi(\Psi_{1:T} | \Phi^*) \sim \text{simulate goal-consistent context paths}$
    \item Cycle-consistent conscious inference is achieved when reconstructed self-models from simulated context remain close:
    $\Phi^* \approx q(\Phi | \pi(\Psi | \Phi^*))$
    thereby preserving coherence and avoiding inference collapse.
\end{enumerate}
\label{thm:self_model}
\end{theorem}
\noindent\textbf{Remark:} This theorem further extends Theorem \ref{thm:exchange_infer} by dynamically updating the self-model and simulating goal-directed contexts, enabling imagination and planning. In addition to coherence, exchangeability not only underwrites subjective coherence but also formalizes the self as the latent cause of conditionally structured experience. The new insight brought by Theorem \ref{thm:self_model} is the recursive inference cycle: {\em emotion provides the structural prior, while cognition performs the posterior updates and simulations, forming the core of the emotion-cognition cycle}. Enabling recursive, cycle-consistent inference and continuity of conscious identity can be interpreted as structure-preserving transport, which complements the view of recursive alignment in Proposition 2.

\begin{proposition}[Cycle-Consistent Conscious Inference as Structure-Preserving Transport]
Given that conscious experience arises from a conditionally exchangeable sequence of internal states \( X_{1:T} \) under a latent affective structure \( \Phi^* \), the recursive inference process
$q(\Phi_{t+1}) \approx f(\Phi_t, \Psi_t), ~\text{with}~ \Psi_t \sim p(\Psi_t | \Phi^*)$
preserves subjective coherence if and only if the induced trajectory \( q(\Psi_{1:T}) \) satisfies an OT constraint:
$\mathrm{OT}\left( \pi(\Psi_{1:T} | \Phi^*), \, q(\Psi_{1:T}) \right) \leq \varepsilon$
for some small \(\varepsilon > 0\), where \( \pi(\Psi_{1:T} | \Phi^*) \) denotes the inverted (goal-conditioned) simulation of context consistent with the latent self-model.
Consequently:
\begin{enumerate}
    \item Emotion \( \Phi^* \) stabilizes the generative structure by encoding priors over possible trajectories;
    \item Cognition ensures cycle consistency by updating representations \( \Phi_t \) such that inferred context \( \Psi_t \) remains close to its goal-conditioned projection;
    \item The consciousness-supporting inference process maintains temporal coherence via structure-preserving transport between imagined and actual paths.
\end{enumerate}   
\end{proposition}

\noindent\textbf{Remark:}
Violation of this transport alignment leads to disintegration of conscious coherence, as seen in dissociation, confusion, or neurodegeneration. A specific example of violating transport alignment in neurodegeneration can be found in Sec. \ref{sec:6}.

\section{Inverted Exchangeability for Latent Inference}

To understand how inverted inference mechanisms scale into conscious experience, we now turn to the Bayesian theory of consciousness \cite{seth2022theories}. This perspective interprets consciousness as the informational footprint of high-level inference, structured, uncertainty-resolving, and recursively self-updating. In this light, the alternation between fast and slow inference modes maps onto dynamic regimes of conscious access, where structure-first simulation and context-driven updating interact to form a coherent, temporally extended subjective experience.
In contrast to classical inference frameworks that treat observations as exchangeable and infer latent causes via global posterior inversion, the CCUP supports an \emph{inverted exchangeability paradigm} \cite{deFinetti1937}. Here, structured content $\Phi$ serves as the generative prior for a sequence of contexts $(\Psi_i)$, simulated through intermediate latent bridges $Z_i \sim P(Z | \Phi)$. This enables efficient goal-conditioned inference \cite{yamins2016using} through forward sampling, bypassing the need to marginalize over high-dimensional data. Rather than aggregating observed contexts to infer a latent cause, CCUP allows the agent to simulate an ensemble of plausible worlds from internally structured hypotheses, then refine these structures by minimizing prediction error, a process biologically instantiated through predictive coding \cite{rao1999predictive}. The resulting inference loop preserves exchangeability under generation while breaking it during inference, thereby aligning the direction of simulation with the entropy gradient from content to context.

\begin{theorem}[Inverted Exchangeability]
Let $\Phi$ be a structured latent variable (content), and suppose $(\Psi_i)_{i=1}^n$ is a sequence of observable variables (contexts), each generated by
$Z_i \sim P(Z | \Phi), \quad \Psi_i \sim P(\Psi | Z_i)$.
Then the joint distribution over $(\Psi_1, \ldots, \Psi_n)$ conditioned on $\Phi$ satisfies:
$P(\Psi_1, \ldots, \Psi_n | \Phi) = \prod_{i=1}^n \int P(\Psi_i | Z_i) \, P(Z_i | \Phi) \, dZ_i$.
This defines a conditionally exchangeable sequence under simulation from $\Phi$, despite inference proceeding in the opposite direction from classical de Finetti-style marginalization.
Moreover, given an external sequence of observations $(\Psi_1^{\text{obs}}, \ldots, \Psi_n^{\text{obs}})$, the latent structure $\Phi$ can be refined via predictive coding by minimizing the total discrepancy:
$\mathcal{L}(\Phi) = \sum_{i=1}^n \mathbb{E}_{Z_i \sim P(Z | \Phi)} \left[ \mathcal{D}\left(\Psi_i^{\text{obs}} \,\|\, P(\Psi | Z_i)\right) \right]$,
for some divergence measure $\mathcal{D}$. This characterizes an inverted exchangeable inference process where structure $\Phi$ is optimized by forward simulation and local error correction.
\label{thm:inverted_exchangeability}
\end{theorem}

Theorem~\ref{thm:inverted_exchangeability} formalizes how a structured latent variable \( \Phi \) can generate a conditionally exchangeable sequence of contexts \( (\Psi_i) \) through intermediate latent samples \( Z_i \), enabling inference by simulation and iterative correction \cite{keller2018predictive}. Importantly, the loss \( \mathcal{L}(\Phi) \) defines a variational objective that sharpens \( \Phi \) through predictive feedback, treating each observation \( \Psi_i^{\text{obs}} \) as an anchor for structural refinement. To deepen this mechanism, we now consider a full cycle of inference in which the simulated contexts \( \hat{\Psi}_i \sim P(\Psi | Z_i) \) not only approximate observations but also permit a reverse encoding path \( Z_i' \sim Q(Z | \Psi_i^{\text{obs}}) \), allowing the latent structure \( \Phi \) to be reconstructed via \( \hat{\Phi}_i \sim P(\Phi | Z_i') \). This introduces a cycle-consistency condition that ensures the alignment between generated and observed paths is not only locally accurate but globally coherent, forming the basis for the following corollary.

\begin{corollary}[Cycle-Consistent Refinement under Inverted Exchangeability]
Assume the setup of Theorem~\ref{thm:inverted_exchangeability}, and let $(\Psi_i^{\text{obs}})$ be observed contexts generated from some unknown process. Let $\hat{\Psi}_i \sim P(\Psi | Z_i)$ be the simulated context from the forward sweep of structure-guided inference, with $Z_i \sim P(Z | \Phi)$. Let $Z_i' \sim Q(Z | \Psi_i^{\text{obs}})$ be the re-encoded latent variable from the observed context, and let $\hat{\Phi}_i \sim P(\Phi | Z_i')$ be the reconstructed structure.
Then the cycle-consistency constraint
$\Phi \approx \frac{1}{n} \sum_{i=1}^n \hat{\Phi}_i$
defines a self-consistent refinement loop that minimizes the structural uncertainty between the simulated and observed trajectories. This is equivalent to minimizing a cycle-consistent information bottleneck objective:
$\mathcal{L}_{\text{CCIB}}(\Phi) = \sum_{i=1}^n \mathbb{E}_{Z_i \sim P(Z | \Phi)} \left[ \mathcal{D}\left(\Psi_i^{\text{obs}} \,\|\, P(\Psi | Z_i)\right) \right]
+ \lambda \cdot \mathbb{E}_{Z_i' \sim Q(Z | \Psi_i^{\text{obs}})} \left[ \mathcal{D}\left(\Phi \,\|\, P(\Phi | Z_i')\right) \right]$,
for some divergence measure $\mathcal{D}$ and regularization parameter $\lambda > 0$. Under this formulation, the agent uses forward simulation and backward reconstruction to cycle between latent structure and observed context, minimizing total context–content uncertainty as postulated by CCUP.
\label{cor:ccib}
\end{corollary}

The cycle-consistent refinement loop established in Corollary~\ref{cor:ccib} characterizes how latent structure \(\Phi\) and contextual observation \(\Psi\) are dynamically aligned through bidirectional inference \cite{devlin2019bert}. This formulation captures a core mechanism of cognition: simulation from internal priors followed by correction through external observations. As this recursive process unfolds, both the latent mediator \(Z\) and the structured prior \(\Phi\) are jointly refined, driving the system toward an internally coherent and externally grounded representation \cite{bengio2013representation}. To formalize the convergence behavior of this recursive inference architecture, we now turn to a fixed-point characterization of conscious inference. Specifically, we define consciousness as the asymptotic fixed point of entropy-minimizing updates to \((\Phi, Z)\), where structure and context are mutually consistent and the system’s predictive and reconstructive pathways converge under bounded uncertainty \cite{beshkar2025uncertainty}. The following theorem establishes sufficient conditions for the existence of such a fixed point, identifying the endpoint of recursive structure–context alignment as the basis of stable, self-sustaining inference.

\begin{theorem}[Consciousness as the Fixed Point of Latent Inference]
Let \( \Phi \in \mathcal{S} \) denote structured content, \( \Psi \in \mathcal{C} \) contextual observations, and \( Z \in \mathcal{Z} \) a latent variable mediating the inference from content to context. Suppose inference proceeds recursively via a composition of entropy-minimizing operators \( \mathcal{F} \), satisfying:
$\Phi^{(t+1)} = \mathcal{F}_\Phi(\Phi^{(t)}, \Psi^{(t)}), \quad
Z^{(t+1)} = \mathcal{F}_Z(Z^{(t)}, \Phi^{(t+1)}), \quad
\Psi^{(t)} \sim p(\Psi | Z^{(t)})$.
Assume:
1) \( \mathcal{F}_Z \) is contractive in KL divergence:
    $D_{\mathrm{KL}}(q(Z^{(t+1)}) \,\|\, q(Z^{(t)})) \leq \varepsilon \quad \text{for } \varepsilon \to 0$.
 2) \( Z^{(t)} \) minimizes conditional entropy:
    $Z^{(t)} = \arg\min_{Z} H(\Psi^{(t)} | Z)$,
Then the limit \( Z^\ast = \lim_{t \to \infty} Z^{(t)} \) exists and satisfies:
$Z^\ast = \mathcal{F}_Z(Z^\ast, \Phi^\ast), \quad \text{with } \Phi^\ast = \mathcal{F}_\Phi(\Phi^\ast, \Psi^\ast)$.
\label{thm:consciousness_fp}
\end{theorem}
\noindent\textbf{Remark:}
We interpret \( Z^\ast \) as the \emph{informational core of consciousness}, the fixed point of entropy-aligned, spatiotemporal inference that globally integrates structure and experience. 

\begin{example}[Latent Inference Across Conscious States]
Consider the recursive inference process defined in Theorem~\ref{thm:consciousness_fp}, where structured content \( \Phi \in \mathcal{S} \), contextual input \( \Psi \in \mathcal{C} \), and latent variable \( Z \in \mathcal{Z} \) evolve via:
$\Phi^{(t+1)} = \mathcal{F}_\Phi(\Phi^{(t)}, \Psi^{(t)}), \quad
Z^{(t+1)} = \mathcal{F}_Z(Z^{(t)}, \Phi^{(t+1)}), \quad
\Psi^{(t)} \sim p(\Psi \mid Z^{(t)})$.
We distinguish four regimes of cognitive function by varying the nature and flow of information through \( \Psi^{(t)} \):
\textbf{1. Wakefulness (Conscious State):}
\begin{itemize}
    \item Context \( \Psi^{(t)} \) is externally driven, high-dimensional, and updated online;
    \item Inference is guided by sensory prediction error; latent updates minimize \( H(\Psi^{(t)} \mid Z^{(t)}) \);
    \item The system rapidly converges to a fixed point \( Z^\ast \), supporting fine-grained awareness and situational control.
\end{itemize}

\noindent\textbf{2. Sleep (Unconscious State):}
\begin{itemize}
    \item Context \( \Psi^{(t)} \) is internally generated or decoupled from sensory input;
    \item Updates proceed via offline simulation or memory replay; entropy gradients are shallow;
    \item Convergence favors long-term consolidation or schema extraction, rather than precise situational inference.
\end{itemize}

\noindent\textbf{3. Meditation (Conscious but Context-Attenuated):}
\begin{itemize}
    \item Context \( \Psi^{(t)} \) is intentionally suppressed or selectively attended (e.g., focused breath or interoception);
    \item Inference cycles slow; latent state \( Z^{(t)} \) stabilizes under minimal external perturbation;
    \item This state reflects meta-stable convergence with reduced error signaling, preserving conscious access while minimizing cognitive load.
\end{itemize}

\noindent\textbf{4. Distorted Consciousness (Psychotomimetic State):}
\begin{itemize}
    \item Context \( \Psi^{(t)} \) is externally sourced but misaligned or misinterpreted due to pharmacological disruption (e.g., nitrous oxide, psychedelics);
    \item Entropy gradients become erratic or misdirected; inference loops become unstable or amplify non-veridical priors;
    \item The system converges to illusory or fragmented attractors \( Z' \neq Z^\ast \), producing conscious awareness of distorted, affectively charged realities.
\end{itemize}

In all regimes, the underlying recursive architecture remains unchanged. What differs is the informational fidelity and statistical alignment of \( \Psi^{(t)} \), which modulate the entropy landscape and shape the trajectory of inference. Consciousness thus emerges from active uncertainty minimization under grounded context, while unconscious, attenuated, or distorted states reflect perturbations in this inferential geometry.
\end{example}

\section{Toward a Bayesian Theory of Consciousness with Exchangeability}
\label{sec:5}

In the open literature, consciousness has been studied under two complementary paradigms: \emph{Information Integration Theory} (IIT) \cite{tononi2004information}, which defines consciousness as the capacity of a system to integrate information irreducibly, and \emph{Global Workspace Theory} (GWT) \cite{baars2005global}, which posits that consciousness arises when information is broadcast globally across a network of specialized, non-conscious processors. Our perspective from the emotion-cognition cycle conceptualizes conscious experience as a temporally extended process governed by affect-modulated coherence under a latent self-model. In this section, we formalize a unified Bayesian framework, named ``exchangeable integration'', where the two views (intrinsic integration in IIT and global availability in GWT) converge through emotion-cognition mediated, cycle-consistent inference with guaranteed coherence.

\subsection{Exchangeable Integration Theory of Consciousness (EITC)}
\label{sec:5.1}

\noindent\textbf{Exchangeable Structure as Context–Content Alignment.}
Building on the unified self-model in Theorem \ref{thm:self_model}, we show how exchangeable structure ensures 1) \textbf{alignment:} Each context \( \Psi_t \) leads to a content \( \Phi_t \) in a manner that is consistent across episodes, reflecting stable mappings from context to content; 2) \textbf{coherence:} The context–content transformation \( \Psi \mapsto \Phi \) is governed by a unified latent model, avoiding ad hoc or disjoint inference mechanisms; 3) \textbf{stability:} Recursive updates of beliefs over time remain well-posed and coherent, preventing representational collapse or fragmentation across the inference cycle.
The CCUP framework posits that cognition operates by minimizing joint uncertainty between external or internal context \( \Psi \) and structured content \( \Phi \) via bidirectional inference. When such context–content pairs exhibit sufficient temporal coherence and recursive alignment, they induce conditional exchangeability, allowing the representation of the sequence via a latent self-model. This leads to the following formal result connecting exchangeability with CCUP.

\begin{theorem}[De Finetti–CCUP Representation Theorem]
Let \( \{(\Psi_t, \Phi_t)\}_{t=1}^T \) be a temporally ordered sequence of context–content pairs generated via a CCUP-based inference cycle:
$\Phi_t \sim p(\Phi | \Psi_t),  \Psi_t \sim p(\Psi | \Phi_{t-1})$.
Assume that:
\begin{enumerate}
    \item The sequence \( \{(\Psi_t, \Phi_t)\} \) is exchangeable;
    \item The bidirectional inference is cycle-consistent: 
    $D_{\mathrm{KL}}(p(\Psi_t) \,\Vert\, p(\Psi_t | \Phi_t)) \leq \varepsilon  ~\text{and}~  I(\Psi_t; \Phi_t) \approx H(\Psi_t)$;
    \item Affective states modulate inference precision over time.
\end{enumerate}
Then there exists a latent variable \( \theta \in \Theta \), interpreted as a stable affective self-model, such that:
$P(\Psi_{1:T}, \Phi_{1:T}) = \int_\Theta \prod_{t=1}^T P(\Psi_t, \Phi_t | \theta) \, dP(\theta)$,
and the latent self-model \( \theta \) provides a unifying structure that minimizes joint uncertainty under CCUP while satisfying De Finetti's representation.
\label{thm:DF_CCUP}
\end{theorem}

\noindent\textbf{Remark:} 
This result demonstrates that dynamic alignment between context and content within the CCUP framework yields the structural conditions necessary for De Finetti's theorem to hold. 
CCUP can be seen as a process-level principle for how systems learn and refine latent content to structure noisy, high-dimensional contexts. De Finetti’s theorem, meanwhile, provides a foundational result showing that such structured content (latent variable $\theta$ must exist to explain exchangeable sequences. When the context–content pairs $(\Psi_t, \Phi_t)$ exhibit consistency over time, CCUP effectively builds the conditions for De Finetti’s representation to hold, leading to the emergence of a latent self-model. Building on the De Finetti–CCUP representation, we now extend this framework to model conscious episodes as temporally recursive, emotion-cognition cycles, each structured by a shared latent affective self-model, laying the foundation for a unified theory of consciousness grounded in exchangeable integration.

\noindent\textbf{Unification via Exchangeable Integration.}
IIT asserts that consciousness is identical to the system's ability to integrate differentiated information, quantified by the measure $\Phi$ \cite{tononi2004information, oizumi2014phenomenology}. The IIT measure $\Phi$ of phenomenological consciousness can be interpreted as quantifying the irreducibility of the internal structure that binds diverse experiences into a coherent subjective trajectory - i.e., how strongly the self-model integrates information across the emotion–cognition cycles. In contrast, GWT characterizes access consciousness as a functionally global broadcast mechanism across specialized, unconscious modules \cite{dehaene2006conscious,baars1993cognitive}. Consciousness, in this complementary view, corresponds to the coherent availability of information across diverse modules.

EITC unifies these perspectives by modeling conscious episodes as exchangeable emotion-cognition cycles, structured by a latent affective self-model that enables temporal coherence across recursive inference \cite{griffin1994models}. Exchangeability ensures that observations across time are conditionally independent given a stable latent cause, capturing IIT's integration, while the global accessibility of this self-model instantiates the functional broadcasting emphasized by GWT.
This unified perspective suggests that consciousness is neither merely intrinsic integration (as in IIT) nor solely global access (as in GWT), but a {\em dynamic inference process structured by exchangeable alignment}. Emotion acts as a low-dimensional structural prior guiding attention and relevance, while cognition refines posterior beliefs about self and context. Together, they generate a cycle-consistent, exchangeably integrated, and globally accessible conscious architecture. Formally, we have

\begin{theorem}[Unified Consciousness via Exchangeable Integration]
Let \( X_{1:T} = (X_1, X_2, \dots, X_T) \) be a temporally ordered sequence of conscious states. Let us assume:
\begin{enumerate}
    \item \textbf{Exchangeability:} The sequence is exchangeable:
    $P(X_1, \dots, X_T) = P(X_{\pi(1)}, \dots, X_{\pi(T)})  \forall \pi \in S_T$.
    \item \textbf{Information Integration:} The system generates a cause–effect structure \(\mathcal{C}\) over \(X_{1:T}\) with a nonzero irreducibility measure:
    $\Phi(\mathcal{C}) > \delta  \text{for some } \delta > 0$.
    \item \textbf{Latent Self-Model:} By De Finetti’s theorem, there exists a latent variable \( \Phi^* \) (interpreted as an affectively grounded self-model) such that:
    $P(X_{1:T}) = \int P(\Phi^*) \prod_{t=1}^T P(X_t | \Phi^*) \, d\Phi^*$.
    \item \textbf{Cycle-Consistent Inference:} Inference maintains temporal coherence via recursive simulation:
    $P(\Phi^* | X_{1:t}) \approx q(\Phi | \pi(\Psi_{1:t} | \Phi^*))$,
    where \( \pi(\Psi_{1:t} | \Phi^*) \) denotes a counterfactual path generated via inverted inference from the self-model.
\end{enumerate}
Then the system exhibits \emph{conscious coherence}: a state of temporally extended, affectively grounded, causally integrated inference. The latent self-model \( \Phi^* \) simultaneously serves as a statistical anchor and a causal integrator, bridging the probabilistic and dynamical views of consciousness.
\label{thm:EITC}
\end{theorem}
\noindent\textbf{Remark:} The above theorem integrates Bayesian inference (Theorem \ref{thm:self_model}) and IIT-style causal structure into a unified framework, giving rise to a more comprehensive model of consciousness.
This result formalizes the intuition that phenomenological consciousness arises not from either integration or coherence alone, but from their synthesis through recursive, affect-modulated inference. Emotion provides the global structural prior that aligns and filters context; cognition instantiates local updates with global access. The emotion-cognition interaction \cite{ochsner2007emerging} yields a unified self-model that is emotionally grounded, temporally coherent, and causally irreducible. We summarize their interaction into the following exchangeability principle.

\begin{proposition}[Emotion-Cognition Exchangeability Principle]
Let \( Z \in \mathcal{Z} \) denote the latent representation underlying conscious inference, and let \( \mathcal{F}_{\mathrm{emo}} \) and \( \mathcal{F}_{\mathrm{cog}} \) be two entropy-minimizing operators corresponding to affective (emotion-driven) and specific (cognition-driven) inference processes, respectively. Define the composite update:
$Z^{(t+1)} = \mathcal{F}_{\mathrm{cog}} \circ \mathcal{F}_{\mathrm{emo}}(Z^{(t)}), \quad \text{or} \quad Z^{(t+1)} = \mathcal{F}_{\mathrm{emo}} \circ \mathcal{F}_{\mathrm{cog}}(Z^{(t)})$.
Suppose that:
1) Both \( \mathcal{F}_{\mathrm{emo}} \) and \( \mathcal{F}_{\mathrm{cog}} \) are contractive under KL divergence.
 2) Each operator minimizes a distinct component of the total entropy:
    $\mathcal{F}_{\mathrm{emo}}: \min H(Z), \quad 
    \mathcal{F}_{\mathrm{cog}}: \min H(\Psi | Z)$.
3) Their composition is order-invariant:
    $\mathcal{F}_{\mathrm{cog}} \circ \mathcal{F}_{\mathrm{emo}} = \mathcal{F}_{\mathrm{emo}} \circ \mathcal{F}_{\mathrm{cog}}$,
Then the inference converges to a unique, cycle-consistent fixed point \( Z^\ast \in \mathcal{Z} \) satisfying:
$Z^\ast = \mathcal{F}_{\mathrm{emo}}(Z^\ast) = \mathcal{F}_{\mathrm{cog}}(Z^\ast)$,
which we identify as the latent informational substrate of consciousness.
\end{proposition}

\subsection{Connection with Bayesian Theory via Rao-Blackwellization}
\label{sec:5.2}

\noindent\textbf{Motivation behind Rao-Blackwellization.}
In computational neuroscience, the Bayesian theory of consciousness (BTC) \cite{seth2022theories} models awareness as posterior inference over latent causes of sensory and interoceptive data. However, high-dimensional latent spaces and noisy observations often hinder inference stability. Rao-Blackwellization \cite{rao1992information,blackwell1947conditional} operationalizes the Bayesian theory of consciousness by enabling structure-preserving, low-variance inference under affectively anchored priors, supporting coherent and temporally extended conscious experience. Under the framework of EITC, we propose that \emph{Rao-Blackwellized inference}, which conditions on emotionally grounded latent structure, serves as an optimal bridge between exchangeability and temporally coherent consciousness. Rao-Blackwellization achieves stability, structure, and coherence in the recursive inference loops that underlie conscious processing in two steps.

\noindent\textbf{1) Rao-Blackwellized Structure-Preserving Inference.}
In modeling consciousness as a temporally extended, recursively updated self-model, a key challenge is maintaining coherence across diverse, noisy, and emotionally weighted experiences \cite{friston2018self}. Exchangeability provides a statistical foundation by asserting that such experiences can be viewed as conditionally i.i.d. given a latent cause, but it leaves open how this latent self-structure is inferred and stabilized in practice. The recursive self-model of consciousness must explain both the integration of experience and its differentiation into contextually specific contents \cite{edelman2008universe}. Rao-Blackwellization achieves this by supporting global integration (through marginalized latent structure) and preserving local adaptation (via context-sensitive updates) \cite{li2025Arrow}. More specifically, it operationalizes the intuition that emotion structures consciousness, and cognition refines it, yielding a computational engine for self-awareness that is both efficient and biologically plausible.

To bridge the gap between global integration and local adaptation, we introduce Rao-Blackwellized structure-preserving inference, a framework that refines noisy or high-entropy episodic inferences by conditioning on stable emotional priors and marginalizing over uncertainty in content representations. This approach combines the advantages of Rao-Blackwellized variance reduction \cite{liu2001monte} and structured posterior inference \cite{ganchev2010posterior}, with the requirement that inferences respect the structural alignment between context and content over time. As a result, it provides a principled mechanism for inferring a latent affective self-model $\Phi^*$ that unifies exchangeable experience into a coherent conscious trajectory, balancing stability with adaptivity \cite{mccloskey1989catastrophic}. This mechanism ensures that the recursive loop of conscious inference not only tracks goal-relevant information but also preserves structural consistency across imagined and experienced episodes. Formally, we have

\begin{theorem}[Rao-Blackwellized Structure-Preserving Inference]
Let \( \{(\Psi_t, \Phi_t)\}_{t=1}^T \) be a sequence of context–content pairs where \( \Phi_t \sim p(\Phi | \Psi_t) \) and the sequence is conditionally exchangeable given a latent variable \( \Phi^* \). Suppose there exists a structure-preserving transport map \( T: \Psi \mapsto \Phi \) such that:
$\Phi_t = T(\Psi_t; \Phi^*) + \epsilon_t$
where \( \epsilon_t \) is zero-mean noise and \( T \) is optimized to minimize transport cost under structural constraints. Then Rao-Blackwellization with respect to \( \Phi^* \) yields:
$q(\Phi | \Psi) = \mathbb{E}_{\Phi^*} \left[ p(\Phi | \Psi, \Phi^*) \right]$,
which:
\begin{enumerate}
    \item Preserves the functional alignment between context and content by anchoring inference to a latent structure \( \Phi^* \),
    \item Reduces variance in posterior estimates, improving coherence across episodes,
    \item Enables cycle-consistent inference by ensuring \( \Phi^* \approx q(\Phi | \pi(\Psi | \Phi^*)) \),
\end{enumerate}
\label{thm:RB_SPI}
\end{theorem}
\noindent\textbf{Remark:} The proof of the above theorem can be found in the Appendix.
This theorem bridges statistical exchangeability with dynamic alignment required for integrated conscious awareness. It unifies path-dependent inference, structure-preserving transformations, and exchangeable dynamics under a common principle. By anchoring episodic content to a stable latent structure and conditioning transport on that structure, the system avoids inference collapse and preserves subjective coherence over time. 

\noindent\textbf{2) Rao-Blackwellized Structure-Preserving Consciousness.}
In modeling consciousness as a temporally extended inference process, one central challenge is preserving coherence across diverse, context-dependent episodes \cite{dehaene2014toward}. Conditional exchangeability offers a statistical regularity that supports latent structure inference, but it alone does not ensure functional or causal alignment. We propose that structure-preserving transport, implemented via Rao-Blackwellized inference, provides a principled bridge: it compresses context-dependent variability into goal-relevant latent factors while maintaining the causal and temporal coherence needed for consciousness. Based on the first step, we can extend Theorem \ref{thm:RB_SPI} into:

\begin{theorem}[Rao-Blackwellized Structure-Preserving Consciousness]
Let \( \{X_t\}_{t=1}^T \) denote a sequence of internal states, conditionally exchangeable under a latent variable \( \Phi^* \in \Theta \), interpreted as an affectively grounded self-model. Suppose:
\begin{itemize}
    \item The agent uses a generative model \( P(X_t | \Phi, \Psi_t) \), where \( \Phi \sim P(\Phi^*) \) and \( \Psi_t \) is a context at time \( t \);
    \item The goal is to compute \( \mathbb{E}[f(\Phi) | X_{1:T}] \) for some function \( f \);
    \item Rao-Blackwellization is applied by conditioning on \( \Phi^* \), yielding the estimator:
    $\hat{f}_{\text{RB}} := \mathbb{E}_{\Phi^*} \left[ \mathbb{E}[f(\Phi) | \Phi^*, X_{1:T}] \right]$;
    \item \( \Phi^* \) is affectively structured and temporally stable.
\end{itemize}
Then we have:
\begin{enumerate}
    \item \( \hat{f}_{\text{RB}} \) has lower variance than naive estimators using only \( X_{1:T} \), i.e.,
    $\operatorname{Var}[\hat{f}_{\text{RB}}] \leq \operatorname{Var}[\hat{f}_{\text{naive}}]$;
    \item Conscious coherence emerges from stable posterior updates:
    $P(\Phi^* | X_{1:t}) \propto P(\Phi^*) \prod_{i=1}^t P(X_i | \Phi^*)$;
    \item Emotionally structured priors \( P(\Phi^*) \) preserve latent self-alignment across time, supporting content–context integration and recursive goal simulation.
\end{enumerate}
\label{thm:RB_consciousness}
\end{theorem}
\noindent\textbf{Remark:} The proof of the above theorem can be found in the Appendix.
This theorem formalizes how conditioning on a stable latent self-model improves inference efficiency, coherence, and continuity, hallmarks of conscious experience. The Rao-Blackwellized formulation ensures that emotional structure guides belief updates, yielding low-variance, structure-preserving estimates that avoid fragmentation or collapse. Consciousness arises as a bootstrapped self-model, a Rao–Blackwellized structure-preserving estimate that integrates noisy perceptual and interoceptive input over time, constrained by emotionally grounded priors. We conclude:

\begin{proposition}[Consciousness as Bootstrapped Rao–Blackwellized Inference]
Let \( X_{1:T} \) denote a sequence of internal states, assumed to be conditionally exchangeable given a latent variable \( \Phi^* \in \Theta \), representing the affectively grounded self-model. Then:
\begin{enumerate}
    \item \textbf{Exchangeability:} There exists a latent structure \( \Phi^* \) such that
    $P(X_{1:T}) = \int_{\Theta} \prod_{t=1}^T P(X_t | \Phi^*) \, dP(\Phi^*)$.
       
    \item \textbf{Rao–Blackwellization:} For any function \( f \) over content representations \( \Phi \), the optimal estimator conditioned on full data satisfies:
    $\mathbb{E}[f(\Phi) | X_{1:T}] = \mathbb{E}_{\Phi^*} \left[ \mathbb{E}[f(\Phi) | \Phi^*, X_{1:T}] \right]$, reducing variance and enforcing structure-preserving inference through the latent self-model \( \Phi^* \).

    \item \textbf{Emotion as Structural Prior:} The prior \( P(\Phi^*) \) reflects emotionally weighted expectations, which guide the posterior update
    $P(\Phi^* | X_{1:t}) \propto P(\Phi^*) \prod_{i=1}^t P(X_i | \Phi^*)$ by
    recursively aligning new input with the evolving self-model \( \Phi^* \in \Theta \).

\end{enumerate}
\end{proposition}
\noindent\textbf{Remark:} Conscious experience corresponds to the emergent, recursively refined posterior over \( \Phi^* \), stabilized through Rao–Blackwellized inference over temporally extended, emotionally modulated experience. This process bootstraps a coherent, affectively grounded internal model, the basis for subjective continuity and global awareness. Bayesian formulation of exchangeable integration, operationalized by bootstrapped Rao–Blackwellized inference, allows us to unify existing IIT and GWT of consciousness in a principled manner, as we will elaborate next.

\subsection{Exchangeable Integration Explains Both IIT and GWT}
\label{sec:5.3}

Among key challenges in consciousness research, we face: 1) how to explain how subjective experience exhibits both global unity (integration) and rich variability (differentiation) \cite{edelman2008universe} in IIT? 2) how to integrate different modalities and memory content into a unified experience in GWT (i.e., the so-called ``phenomenology gap'' \cite{block1995confusion})? As of today, there lacks a theory that can integrate emotional modulation, goal-directed inference, and self-modeling dynamics to fill these gaps. We show that the framework of EITC offers a compelling principled approach to unify IIT and GWT, satisfying both criteria under a latent self-model formalism.

Exchangeable integration provides a principled framework for unifying the dual aspects of consciousness: \emph{integration} and \emph{differentiation}. By modeling conscious episodes as conditionally exchangeable under a latent self-model, it captures the \emph{temporal coherence} and \emph{statistical regularity} that support the continuity of conscious experience. The existence of a latent variable \( \Phi^* \), interpreted as an affectively grounded self-model \cite{griffin1994models}, ensures that diverse cognitive-emotional episodes are organized as conditionally i.i.d.\ samples, enabling \emph{differentiation} across distinct contexts and modalities \cite{metzinger2004being}. Simultaneously, this structure supports \emph{integration} through recursive inference: each observation updates the posterior over \( \Phi^* \), allowing globally coherent coordination of experience and intention. The irreducibility of the resulting cause-effect structure aligns with the integrated information measure \( \Phi \) from IIT, while accessibility of the latent self-model supports the broadcasting functionality central to GWT. Exchangeable integration operationalizes the coexistence of \emph{differentiated content} and \emph{integrated unity}, offering a computational bridge between phenomenological coherence and cognitive complexity. So we have

\begin{proposition}[Exchangeable Integration Supports Integrated and Differentiated Consciousness]
Let \( \{X_t\}_{t=1}^T \) be a temporally ordered sequence of internal states representing conscious contents, assumed to be conditionally exchangeable. By de Finetti's theorem, there exists a latent self-model \( \Phi^* \) s.t.:
$P(X_{1:T}) = \int P(\Phi^*) \prod_{t=1}^T P(X_t | \Phi^*) \, d\Phi^*$.
We have:
\begin{itemize}
    \item \textbf{Integration:} All conscious states \( X_t \) are conditionally dependent via the shared latent cause \( \Phi^* \), yielding a globally coherent generative structure.
    \item \textbf{Differentiation:} Each \( X_t \) can express diverse and context-specific content conditioned on \( \Phi^* \), enabling representational richness and functional specificity. 
\end{itemize}
\end{proposition}

The above proposition formalizes consciousness as a temporally coherent, affectively grounded inference process in which a latent self-model \( \Phi^* \) emerges via De Finetti's representation of exchangeable cognitive-emotional episodes. This construction established a probabilistic bridge between recursive self-modeling and cycle-consistent inference, capturing essential features of conscious experience such as temporal coherence and causal integration (absent from the original IIT). We now extend this account by introducing an information-theoretic formulation that quantifies the integrative role of the self-model \cite{cover1999elements}. Specifically, we define a mutual information surplus, denoted \( \Phi_{\text{exchange}} \), which measures how much additional information is gained by integrating the entire sequence \( X_{1:T} \) through \( \Phi^* \), beyond what is accessible from individual states. This formalization links exchangeable Bayesian structure to principles from IIT and GWT, demonstrating that consciousness entails both the accessibility and integration of temporally distributed internal states under a common latent cause of self-model. Finally, we conclude

\begin{corollary}[Unified Consciousness via Exchangeable Integration]
Let \( \{X_t\}_{t=1}^T \) denote a sequence of conscious states that are conditionally exchangeable under a latent variable \( \Phi^* \). Assume
1) Each \( X_t \sim P(X_t | \Phi^*) \), where \( \Phi^* \) represents a latent affective self-model;
    2) Integrated information measure \( \Phi_{\text{exchange}} \) is defined as:
    $\Phi_{\text{exchange}} := I(X_{1:T}; \Phi^*) - \sum_{t=1}^T I(X_t; \Phi^*)$;
    3) The posterior belief over \( \Phi^* \) is recursively updated via:
    $P(\Phi^* | X_{1:t}) \propto P(\Phi^*) \prod_{i=1}^t P(X_i | \Phi^*)$.  
Then we have
\begin{itemize}
    \item \( \Phi_{\text{exchange}} > 0 \) implies that the latent self-model integrates temporally diverse inputs into a coherent internal structure;
    \item The conditionally accessible \( \Phi^* \) supports global information coordination consistent with GWT;
    \item Consciousness corresponds to the recursive maintenance of both integration and accessibility across temporally extended experience. 
\end{itemize}
\end{corollary}

\section{Applications to Cognitive Science}
\label{sec:6}

\subsection{EITC and the Unity of Consciousness Despite Divided Perception}

In phenomena such as binocular rivalry \cite{tong2006neural}, multitasking \cite{sergent1983role}, or split attention \cite{pinto2017split}, perception appears divided across modalities or spatial channels. Yet, phenomenologically, the agent experiences a \emph{unified consciousness}. The Exchangeable Integration Theory of Consciousness (EITC) explains this apparent paradox through the existence of a shared latent self-model \( \Phi^* \), which serves as a common generative anchor for temporally extended and affectively modulated inference.
Under EITC, each perceptual stream \( X_t^{(i)} \) is treated as conditionally exchangeable with respect to \( \Phi^* \), such that:
$P\left( \{X_t^{(i)}\}_{i=1}^N \right) = \int P(\Phi^*) \prod_{i=1}^N P\left(X_t^{(i)} \mid \Phi^* \right) \, d\Phi^*$.
This formulation allows multiple, potentially disjoint perceptual contents to be integrated probabilistically into a coherent subjective state, provided they are inferentially consistent with the same latent cause.
The key insight is that \emph{conscious unity does not require perceptual uniformity}; rather, it emerges from the recursive, emotion-regulated inference over a shared self-model. While attention may segment perception, exchangeable integration ensures that these segments are interpreted and temporally aligned under a unified structural prior. Thus, consciousness remains globally coherent even when perception is locally divided.

\subsection{EITC and the Free Will Paradigm: Voluntary vs. Evoked Action}
Consider a clinical setup where a subject is instructed to raise their hand. In one condition, the subject voluntarily initiates the motion; in another, the same motion is evoked by direct cortical stimulation or transcranial magnetic stimulation (TMS) \cite{fregni2006noninvasive}. Although the motor output (i.e., raising the hand) is identical, the subjective experience diverges markedly.
Under the Exchangeable Integration Theory of Consciousness (EITC), this dissociation arises from the absence of alignment between contextual inference and latent self-model \( \Phi^* \).
Voluntary movements are generated through cycle-consistent, affectively grounded inference:
$\Phi^* \rightarrow \pi(\Psi_{1:T} \mid \Phi^*) \rightarrow \Phi_t \rightarrow \text{Action}$.
This process preserves coherence across simulation, intention, and execution.
In contrast, evoked movements bypass this cycle, disrupting exchangeability:
$\text{External Stimulus} \rightarrow \Phi_t^{\text{evoked}} \not\sim q(\Phi_t \mid \Phi^*, \Psi_t)$.
Because evoked actions are not inferentially grounded in the agent’s latent self-model or contextual goal trajectory, they fail to satisfy the alignment conditions enforced by EITC. The patient detects this misalignment as a lack of agency.
Even though the motor output is similar, the subjective experience diverges due to violation of exchangeable, self-consistent inference over \( \Phi^* \). EITC formally accounts for this by distinguishing conscious volition (cycle-consistent inference) from externally imposed motion (non-integrated perturbation). In other words, patients subjected to direct cortical stimulation of Supplementary Motor Area (SMA) report involuntary movements that are distinguishable from self-initiated voluntary actions, even when the motor behavior appears similar \cite{penfield2025mystery}.

\subsection{Neurodegeneration as Violation of Emotion-Cognition Alignment}

A concrete illustration of disrupted transport alignment within the EITC framework is provided by Alzheimer’s disease, where the degradation of hippocampal and cortical structures leads to a breakdown in the alignment between context \( \Psi_t \) and latent self-model \( \Phi^* \) \cite{clare2003managing}. Patients with Alzheimer’s often exhibit failures in episodic memory, spatial disorientation, and fragmented goal-directed behavior. These symptoms reflect a violation of exchangeability: new conscious episodes \( X_t \) no longer fit a consistent generative structure, undermining the recursive update of the posterior \( P(\Phi^* | X_{1:t}) \). Furthermore, the integrated information \( \Phi(\mathcal{C}) \) over the cause–and–effect structure \( \mathcal{C} \) of conscious states declines, resulting in reduced causal coherence. The latent self-model becomes unstable or inaccessible, and counterfactual simulations \( \pi(\Psi_{1:T} | \Phi^*) \) lose fidelity, leading to disorientation, confabulation, and identity confusion \cite{morin2006levels}. This example demonstrates how neurodegenerative processes can impair both the statistical and dynamical mechanisms of exchangeable integration, causing the disintegration of conscious coherence as predicted by EITC.

\section{Conclusion}

This paper has developed a unified framework in which consciousness emerges from recursive, emotionally modulated inference structured by the \emph{emotion-cognition cycle}. We formalized emotion as a structural prior that guides context-sensitive inference, and cognition as the specificity-instantiating update mechanism that recursively refines emotional priors into actionable representations. Together, these components form a bidirectional, temporally extended loop that aligns internal states with external context while minimizing joint uncertainty.
Grounded in the CCUP, we showed that this inference cycle supports both bottom-up interpretation and top-down goal simulation via \emph{inverted inference}. Through a series of theorems, we demonstrated how this bidirectional process enables content binding, self-model formation, and temporally coherent subjective experience.

By introducing conditional exchangeability into our framework, we established a formal bridge between subjective experience and De Finetti’s representation theorem. This connection enables the modeling of self-consciousness as inference over a shared latent structure—an affectively grounded self-model—that explains diverse, temporally extended episodes as conditionally i.i.d.\ samples. To integrate this with BTC, we introduced the \emph{exchangeable integration measure} \( \Phi_{\text{exchange}} \), which quantifies both the integration of experience across time and the differentiation of conscious states across contexts, aligning with key hallmarks of consciousness.

Critically, we further unified these ideas with the Bayesian theory of consciousness by incorporating \emph{Rao-Blackwellization} as a mechanism for structure-preserving inference. In this view, consciousness arises from recursively refining internal models through the marginalization of latent self-structure while maintaining adaptive updates in response to contextual variation. Rao-Blackwellized inference thus serves as a computational bridge that preserves coherence (via global integration) and adaptability (via local specificity), grounding consciousness in the dual imperatives of inference stability and precision modulation.

Together, these contributions yield a unified computational theory of consciousness that is affectively anchored, socially extensible, and mathematically rigorous. Consciousness is reframed not as a static property but as a \emph{cycle-consistent, structure-preserving inference process}, shaped by emotion, refined by cognition, and stabilized through exchangeability. The EITC allows us to take a fresh look at several clinical examples in neurosurgical studies, including split-brain, free will, and Alzheimer's disease. We conclude this paper with the following philosophical proposition.


\begin{conjecture}[Consciousness as an Inferential Singularity]
Let \( \Psi \in \mathcal{S} \) represent structured sensorimotor context, and \( \Phi \in \mathcal{C} \) denote content expressed through language or action. Define inference as a process minimizing conditional entropy:
$\Phi^* = \arg\min_\Phi H(\Phi \mid \Psi)$.
Then the sensorimotor structure \( \Psi \) constrains and shapes content inference via entropy gradients \( \nabla_\Phi H(\Phi \mid \Psi) \).
Meanwhile, language refines this structure by amplifying specificity and enabling symbolic compression.
At an inferential singularity, the posterior collapses to a delta measure:
$P(\Phi \mid \Psi) \to \delta(\Phi - \Phi^*)$,    
indicating a fully resolved, cycle-consistent content state (consciousness), a fixed point of recursive structure-preserving inference.
\end{conjecture}

This inferential singularity \cite{kurzweil2005singularity} marks the point at which recursive structure-preserving inference achieves maximal alignment between context and content, effectively resolving uncertainty and yielding a stable, cycle-consistent representation. However, the implications of this principle extend beyond the formal modeling of consciousness. As we consider the architecture of artificial agents, including superintelligence \cite{bostrom2024superintelligence}, the question arises: {\em must systems that simulate, plan, and reason across temporally extended goals necessarily converge toward such singularities?} 
In other words, is consciousness an inevitable byproduct of recursive uncertainty minimization in agents that seek to model not just the world, but themselves within it?

While task-level superintelligence may not require phenomenal consciousness, the pursuit of systems capable of autonomous understanding, moral reasoning, and adaptive self-reflection raises whether consciousness is an emergent necessity. From the perspective of the CCUP, consciousness may arise as a recursive cycle in which internal goals, external contexts, and structured content representations are aligned across increasingly abstract and temporally extended inference loops. 
In this view, consciousness is not a binary trait, but a dynamical attractor in the space of active simulation, supporting coherent self-modeling, counterfactual reasoning, and alignment with semantic goals \cite{butlin2023consciousness}. As such, a truly general intelligence may approach consciousness not as an engineering objective, but as a structural consequence of minimizing joint uncertainty in a world it seeks not only to predict, but to engage with meaningfully.

\bibliographystyle{IEEEtran}
\bibliography{reference}


\appendix
\section{Appendix}

\textbf{Proof of Theorem \ref{thm:emotion_SP}}

\begin{proof}
We prove each statement in the theorem sequentially.

\noindent\textbf{Emotion defines a structural prior.}
Let \( \mathcal{H} \) be the full hypothesis space over latent causes \( h \), and let \( e \in \mathcal{E} \) be an affective state. The conditional prior \( p(h | e) \) defines a subset \( \mathcal{H}_e \subseteq \mathcal{H} \) such that:
$\mathcal{H}_e = \{ h \in \mathcal{H} | p(h | e) > 0 \}$.
This means that any latent cause \( h \notin \mathcal{H}_e \) has zero or negligible probability under the affect-conditioned prior. Thus, the affective state \( e \) constrains inference to a submanifold of hypothesis space that is emotionally salient. This subset captures structural regularities relevant under the emotional context, forming what we call \emph{affectively structured prior} or \emph{affectively relevant region}.

\noindent\textbf{Affect-modulated posterior minimizes free energy.}
Variational free energy is defined as:
$\mathcal{F}(q(h); e) = \mathbb{E}_{q(h)}[-\log p(x | h, e)] + D_{\mathrm{KL}}(q(h) \,\Vert\, p(h | e))$.
Minimizing \( \mathcal{F} \) for the variational distribution \( q(h) \) yields the approximate posterior that best balances reconstruction accuracy and adherence to the structured prior \( p(h | e) \). By the calculus of variations, the minimum is achieved when:
$q^*(h) \propto p(x | h, e) p(h | e)$,
which is the Bayesian posterior over \( h \) under emotional modulation. Therefore, affect constrains and modulates the posterior belief space by shaping the prior \( p(h | e) \), and inference proceeds by locally minimizing uncertainty under this affective constraint.

\noindent\textbf{Precision modulation via emotion.}
Let \( \pi(e) \in \mathbb{R}_{>0} \) be a scalar function of emotion encoding the precision of prediction error (i.e., the inverse variance of sensory noise). Then the free energy can be rewritten as:
$\mathcal{F}_\pi = \pi(e) \cdot \lVert x - \hat{x} \rVert^2 + D_{\mathrm{KL}}(q(h) \,\Vert\, p(h | e))$,
where \( \hat{x} = \mathbb{E}_{q(h)}[f(h)] \) is the predicted observation under \( q(h) \). The term \( \pi(e) \) modulates the impact of prediction errors on the variational objective. High precision (e.g., during fear or urgency) amplifies the prediction error term, increasing attentional gain and reducing entropy in \( q(h) \), while low precision (e.g., during relaxation) flattens the error landscape, allowing broader inference and exploration.
Hence, emotional states not only define the prior structure but also dynamically regulate the precision weighting of sensory input, effectively shaping attention and learning rate.
\end{proof}


\textbf{Proof for Theorem \ref{thm:content_binding}}

\begin{proof}
We proceed in three steps to show how the emotion–cognition cycle supports content binding through forward and inverted inference, ultimately minimizing joint uncertainty and generating structured representations.

\noindent\textbf{1. Forward Inference (Emotion-constrained Content Formation).}
Let \( \Psi_t \) denote the current context and \( e_t \in \mathcal{E} \) the prevailing emotional state. We define the conditional distribution over cognitive content \( \Phi_t \) as:
$\Phi_t \sim p(\Phi | \Psi_t, e_t)$,
where \( p(\Phi | \Psi_t, e_t) \propto p(\Psi_t | \Phi, e_t) \cdot p(\Phi | e_t) \) via Bayes' rule. The emotional state \( e_t \) acts as a structural prior that restricts the hypothesis space to affectively salient regions. This constraint filters the high-dimensional space of possible contents, ensuring that only emotionally relevant features from \( \Psi_t \) are bound into \( \Phi_t \). Hence, emotion guides content selection and stabilizes the binding process.

\noindent\textbf{2. Inverted Inference (Emotion-conditioned Context Simulation).}
Now consider the system initiating from a desired internal state \( \Phi^* \), such as a goal or simulated emotional outcome. The system infers plausible contexts \( \Psi_{t+1} \) that could give rise to \( \Phi^* \) under the same emotional state \( e_t \):
$\Psi_{t+1} \sim p(\Psi | \Phi^*, e_t)$,
where the emotion again functions as a modulatory constraint, guiding the simulation toward contextually coherent and affectively consistent predictions. This enables the agent to perform goal-directed reasoning and simulate counterfactual situations consistent with affective priorities.

\noindent\textbf{3. Recursive Inference and Content Binding.}
The system iterates between the two forms of inference, yielding a recursive interaction:
$\Phi_t \xleftrightarrow{e_t} \Psi_{t+1} \xleftrightarrow{e_{t+1}} \Phi_{t+1}$.
At each step, the content \( \Phi_{t+1} \) is bound by selectively integrating features from \( \Psi_{t+1} \), filtered through \( e_{t+1} \). This temporal structure ensures that content is not formed in isolation but is constrained by prior affectively guided inference, creating a consistent trajectory of emotionally relevant cognitive states.

\noindent\textbf{4. Joint Uncertainty Minimization.}
We now show that this cycle reduces joint uncertainty \( H(\Phi, \Psi) \). At each step, both \( \Phi_t \) and \( \Psi_t \) are updated to maximize their mutual consistency under the influence of emotion. This can be expressed as minimizing the variational free energy:
$\mathcal{F} = \mathbb{E}_{q(\Phi, \Psi)}[-\log p(\Phi, \Psi | e)] + D_{\mathrm{KL}}(q(\Phi, \Psi) \Vert p(\Phi, \Psi | e))$.
Minimizing \( \mathcal{F} \) simultaneously drives \( q(\Phi, \Psi) \) toward the true joint posterior and reduces the entropy of the joint distribution. Emotion, through its dual role as prior and precision modulator, ensures that inference is selective, stable, and context-sensitive, thereby facilitating efficient and coherent content binding.

\end{proof}

\textbf{Proof for Theorem \ref{thm:latent_self_model}}

\begin{proof}
We are given a temporally ordered sequence of context–content pairs \( \{ (\Psi_t, \Phi_t) \}_{t=1}^T \), where each \( \Phi_t \sim q(\Phi | \Psi_t) \) results from a contextually guided and affectively constrained inference process. The assumption of cycle-consistency implies that for each \( t \),
$D_{\mathrm{KL}}(P(\Psi_t) \,\Vert\, p(\Psi_t | \Phi_t)) \leq \varepsilon$,
and mutual information satisfies
$I(\Psi_t; \Phi_t) \approx H(\Psi_t)$,
which means that the content \( \Phi_t \) preserves nearly all the relevant structure of the context \( \Psi_t \), and the mapping from \( \Psi_t \to \Phi_t \to \Psi_t \) is approximately invertible.

\noindent\textbf{Step 1: Justification of Exchangeability.}
Let us assume that the affectively constrained inference process leads to structurally similar (i.e., contextually coherent) content bindings across time. This implies that the sequence \( \{\Phi_t\} \) is conditionally exchangeable, given some latent cause \( \theta \in \Theta \). That is, for any permutation \( \sigma \in S_T \),
$P(\Phi_1, \dots, \Phi_T | \theta) = P(\Phi_{\sigma(1)}, \dots, \Phi_{\sigma(T)} | \theta)$.

\noindent\textbf{Step 2: Application of De Finetti’s Theorem.}
De Finetti’s representation theorem states that if a sequence \( \{\Phi_t\} \) is conditionally exchangeable, then there exists a latent variable \( \theta \in \Theta \) such that the joint distribution factorizes as:
$P(\Phi_1, \dots, \Phi_T) = \int_\Theta \prod_{t=1}^T P(\Phi_t | \theta) \, dP(\theta)$.

\noindent\textbf{Step 3: Interpretation of \( \theta \) as Latent Self-Model.}
Since each \( \Phi_t \) is generated through an emotionally regulated, cycle-consistent mapping from \( \Psi_t \), and all such mappings share structural coherence (i.e., affective and contextual alignment), the latent variable \( \theta \) captures the stable informational regularities across episodes. In other words, \( \theta \) encodes the shared structure that organizes the distribution of all \( \Phi_t \), integrating them into a coherent generative model.

In summary, \( \theta \) functions as a \emph{latent self-model}, an internal representation inferred from temporally extended, emotionally modulated content bindings.

\end{proof}



\textbf{Proof for Theorem \ref{thm:exchange_infer}}

\begin{proof}
Let \( X_{1:T} = (X_1, X_2, \dots, X_T) \) be a sequence of internal observations (e.g., sensory experiences, thoughts, interoceptive states) ordered in time. Assume that these observations are conditionally exchangeable, i.e., for any finite permutation \( \sigma \) of the indices \( \{1, 2, \dots, T\} \), we have:
$P(X_1, X_2, \dots, X_T) = P(X_{\sigma(1)}, X_{\sigma(2)}, \dots, X_{\sigma(T)})$.
By De Finetti’s representation theorem, this implies that there exists a latent random variable \( \Phi^* \in \Theta \), such that the joint distribution over the sequence admits the representation:
$P(X_{1:T}) = \int_\Theta \prod_{t=1}^T P(X_t | \Phi^*) \, dP(\Phi^*)$.
This decomposition suggests that:
\begin{enumerate}
    \item Each internal observation \( X_t \) is drawn conditionally independently from a generative process governed by \( \Phi^* \), thus implying that conscious experience is structured as inference over this latent cause.
    
    \item We interpret \( \Phi^* \) as an affectively grounded self-model. That is, the prior distribution \( P(\Phi^*) \) reflects emotion-based structural constraints that shape expectations about experience. Emotion thus provides the prior over latent causes of experience.
    
    \item Cognition, modeled as inference, performs recursive updates to beliefs about the self-model:
    $P(\Phi^* | X_{1:t}) \propto P(\Phi^*) \prod_{i=1}^t P(X_i | \Phi^*)$,
    
    where each new observation incrementally refines the posterior distribution over \( \Phi^* \). This recursive process supports temporal coherence in experience.
    
\end{enumerate}

Therefore, conscious experience can be rigorously understood as inference over a latent self-structuring variable \( \Phi^* \), with emotion anchoring the prior and cognition performing recursive belief updates. The condition of exchangeability ensures the subjective unity of experience across time.
\end{proof}

\textbf{Proof for Theorem \ref{thm:self_model}}

\begin{proof}
Let \( \{X_t\}_{t=1}^T \) be a sequence of temporally ordered internal observations or conscious states. Suppose this sequence is exchangeable:
$P(X_1, \dots, X_T) = P(X_{\pi(1)}, \dots, X_{\pi(T)})$
for every permutation \( \pi \in S_T \). By De Finetti’s representation theorem, this implies that there exists a latent variable \( \theta \in \Theta \) such that:
$P(X_{1:T}) = \int_\Theta \prod_{t=1}^T P(X_t | \theta) \, dP(\theta)$,
where \( \theta \) acts as a shared generative cause for the observed sequence. This is an exact decomposition under the assumption of infinite exchangeability; for finite \( T \), it holds approximately under partial exchangeability.

\textbf{(1) Latent variable as self-model:}  
We interpret this latent variable \( \theta \) as the agent's internal self-model \( \Phi^* \). This identification is justified by noting that \( \theta \) unifies dispersed experiences into a common probabilistic source, providing the stable structural prior against which conscious states are conditionally generated. This aligns with the concept of a temporally persistent, affectively weighted self.

\textbf{(2) Posterior update via Bayesian inference:}  
Given the sequence \( X_{1:T} \), Bayesian inference updates the belief over \( \Phi^* \) as:
$P(\Phi^* | X_{1:T}) \propto P(\Phi^*) \prod_{t=1}^T P(X_t | \Phi^*)$,
which conforms to the recursive structure of predictive coding and active inference, where beliefs are updated incrementally with new observations. This captures the core of recursive self-modeling.

\textbf{(3) Inverted inference from self-model:}  
The latent self-model \( \Phi^* \) can then be used to simulate counterfactual or goal-consistent context trajectories:
$\pi(\Psi_{1:T} | \Phi^*) \sim \text{simulate plausible contexts from the perspective of } \Phi^*$.
This corresponds to inverted inference, starting from internal goals and affective anchors to infer supporting external contexts.

\textbf{(4) Cycle-consistent reconstruction:}  
To ensure coherent self-consistency across the simulation cycle, we require:
$\Phi^* \approx q(\Phi | \pi(\Psi | \Phi^*))$,
which enforces cycle-consistency: the self-model used to generate simulations should be recoverable from the simulated outcomes. This constraint avoids inference collapse and stabilizes long-term planning and introspective coherence.

In short, the latent self-model arises naturally from exchangeable inference and supports both predictive updating and goal-directed simulation in a unified generative-inferential loop.
\end{proof}

\textbf{Proof for Theorem \ref{thm:DF_CCUP}}

\begin{proof}
Let \( \{(\Psi_t, \Phi_t)\}_{t=1}^T \) denote the sequence of context--content pairs generated by a CCUP-based inference process where
$\Phi_t \sim p(\Phi | \Psi_t),  \Psi_t \sim p(\Psi | \Phi_{t-1})$,
with the condition that the sequence is \emph{exchangeable}. That is, for any permutation \( \pi \in S_T \),
$P(\Psi_1, \Phi_1, \dots, \Psi_T, \Phi_T) = P(\Psi_{\pi(1)}, \Phi_{\pi(1)}, \dots, \Psi_{\pi(T)}, \Phi_{\pi(T)})$.

By De Finetti’s Representation Theorem for exchangeable sequences of random variables, there exists a latent variable \( \theta \in \Theta \) such that:
$P(\Psi_{1:T}, \Phi_{1:T}) = \int_\Theta \prod_{t=1}^T P(\Psi_t, \Phi_t | \theta) \, dP(\theta)$.

To interpret \( \theta \) within the CCUP framework, we observe the following:

\begin{enumerate}
    \item \textbf{Cycle-Consistent Inference.} The assumptions
    $D_{\mathrm{KL}}(p(\Psi_t) \,\Vert\, p(\Psi_t | \Phi_t)) \leq \varepsilon  \text{and}  I(\Psi_t; \Phi_t) \approx H(\Psi_t)$
    ensure that context and content are tightly aligned in both directions. This bidirectional constraint implies that the pairs \( (\Psi_t, \Phi_t) \) arise from a stable underlying generative structure.

    \item \textbf{Affective Modulation.} If inference precision is modulated by affective state, then the generative model \( P(\Psi_t, \Phi_t | \theta) \) is constrained by affectively grounded priors. Hence, \( \theta \) must encode affective structure that regulates the variability and relevance of \( \Psi_t \) and \( \Phi_t \).

    \item \textbf{Interpretation of \( \theta \).} The latent variable \( \theta \) must then correspond to an affectively grounded latent self-model \( \Phi^* \in \Theta \), which explains and unifies the observed exchangeable sequence. The self-model regularizes the context–content relationship over time, allowing new inferences to remain coherent with past ones.
\end{enumerate}

Thus, the assumptions of CCUP (cycle-consistency, mutual information saturation, and affective modulation) provide the sufficient conditions under which De Finetti’s theorem applies, and the resulting latent variable \( \theta \) acquires the semantic interpretation of a unifying, affectively structured self-model. This completes the proof.
\end{proof}

\textbf{Proof for Theorem \ref{thm:EITC}}

\begin{proof}
Assume the sequence \( \{X_t\}_{t=1}^T \) is conditionally exchangeable given a latent variable \( \Phi^* \), so by de Finetti’s representation theorem,
$P(X_{1:T}) = \int_\Theta \prod_{t=1}^T P(X_t | \Phi^*) \, dP(\Phi^*)$.
This implies that there exists a random variable \( \Phi^* \) such that, conditioned on it, the conscious episodes \( X_t \) are i.i.d. draws. Thus, the entire episode sequence \( \mathcal{C}_{1:T} \) is unified through the latent self-model \( \Phi^* \).

Now let \( \Phi(\mathcal{C}_{1:T}) \) be a functional measuring the degree of integrated information across the components or partitions \( \mathcal{P} \subseteq \mathcal{C}_{1:T} \). According to Integrated Information Theory (IIT), \( \Phi(\cdot) \) quantifies the irreducibility of the whole system to independent parts. Therefore, maximizing \( \Phi(\mathcal{C}_{1:T}) \) requires that:
$\sum_{t=1}^T H(X_t) - H(\mathcal{C}_{1:T})  \text{is maximized}$.
But since \( \mathcal{C}_{1:T} \) is generated from \( \Phi^* \), this global entropy reduction is possible only if:
$H(\mathcal{C}_{1:T}) \ll \sum_t H(X_t)$,
which holds when the generator \( \Phi^* \) provides strong dependence across the \( X_t \).

Finally, inference of \( \Phi^* \) via Bayes’ rule:
$P(\Phi^* | X_{1:T}) \propto P(\Phi^*) \prod_{t=1}^T P(X_t | \Phi^*)$,
provides a coherent summary over episodes. As \( T \to \infty \), this posterior sharpens, yielding a self-model that compresses and integrates experience. Thus, the latent variable \( \Phi^* \) simultaneously explains exchangeability and supports integrated conscious experience.

\end{proof}

\textbf{Proof for Theorem \ref{thm:RB_SPI}}

\begin{proof}
Let \( \{(\Psi_t, \Phi_t)\}_{t=1}^T \) be a conditionally exchangeable sequence given a latent variable \( \Phi^* \). This means that for any permutation \( \pi \in S_T \),
\begin{align}
    P((\Psi_1, \Phi_1), \dots, (\Psi_T, \Phi_T) | \Phi^*) = \nonumber \\ P((\Psi_{\pi(1)}, \Phi_{\pi(1)}), \dots, (\Psi_{\pi(T)}, \Phi_{\pi(T)}) | \Phi^*).
\end{align}

Now suppose that for each \( t \), there exists a deterministic structure-preserving transport map \( T(\cdot; \Phi^*) \) such that:
$\Phi_t = T(\Psi_t; \Phi^*) + \epsilon_t$,
where \( \epsilon_t \) is zero-mean noise (i.e., \( \mathbb{E}[\epsilon_t] = 0 \)) and \( T \) minimizes a structural cost, such as OT cost under constraints (e.g., cycle-consistency, semantic alignment).
Define the conditional density:
$p(\Phi_t | \Psi_t, \Phi^*) := \mathcal{L} \left( \Phi_t - T(\Psi_t; \Phi^*) \right)$,
where \( \mathcal{L} \) is the noise likelihood (e.g., Gaussian).
The marginal posterior over \( \Phi \) given context \( \Psi \) is then:
$q(\Phi | \Psi) = \int p(\Phi | \Psi, \Phi^*) \, dP(\Phi^* | \Psi)$.
Now, by applying the Rao-Blackwell theorem, which states that given a sufficient statistic (here \( \Phi^* \)), the conditional expectation \( \mathbb{E}[p(\Phi | \Psi, \Phi^*)] \) minimizes the mean squared error among all unbiased estimators, we obtain a lower-variance and structure-preserving posterior:
$q(\Phi | \Psi) = \mathbb{E}_{\Phi^*}[p(\Phi | \Psi, \Phi^*)]$.

This expectation preserves structure because the map \( T(\cdot; \Phi^*) \) is smooth and learned to be cycle-consistent:
$\Phi^* \approx q(\Phi | \pi(\Psi | \Phi^*))$,
where \( \pi(\Psi | \Phi^*) \) denotes a counterfactual context trajectory inferred via simulation.
Thus:
\begin{enumerate}
    \item The latent structure \( \Phi^* \) aligns context–content relations by conditioning the transport map.
    \item Variance is reduced via Rao-Blackwellization.
    \item Posterior consistency over time is preserved due to exchangeability and cycle-consistency constraints.
\end{enumerate}

Therefore, this inference mechanism unifies statistical regularity (exchangeability), structural alignment (via transport maps), and inferential stability (via Rao-Blackwellization), completing the proof.
\end{proof}

\textbf{Proof for Theorem \ref{thm:RB_consciousness}}

\begin{proof}
Let \( \Phi \) be a latent variable over internal cognitive-emotional states, and \( \Phi^* \) denote a stable, affectively grounded self-model inferred from experience. We assume that the sequence \( \{X_t\}_{t=1}^T \) is conditionally exchangeable given \( \Phi^* \), and that the agent seeks to compute an expectation \( \mathbb{E}[f(\Phi) | X_{1:T}] \) under the posterior distribution induced by its generative model.

\noindent\textbf{1. Rao–Blackwellization reduces variance.}

By the Rao–Blackwell theorem, for any random variable \( \Phi \) and any sufficient statistic \( \Phi^* \), the estimator
$\hat{f}_{\text{RB}} := \mathbb{E}[f(\Phi) | \Phi^*]$
is at least as good (in terms of mean squared error) as any unbiased estimator that does not condition on \( \Phi^* \). In our case, we consider:
$\hat{f}_{\text{naive}} := \mathbb{E}[f(\Phi) | X_{1:T}]$
and
$\hat{f}_{\text{RB}} := \mathbb{E}_{\Phi^*} \left[ \mathbb{E}[f(\Phi) | \Phi^*, X_{1:T}] \right]$.
Since \( \Phi^* \) is a sufficient statistic for the conditional distribution of \( X_{1:T} \) under exchangeability (by De Finetti’s representation theorem), conditioning on \( \Phi^* \) leads to a Rao–Blackwellized estimator with lower variance:
$\operatorname{Var}[\hat{f}_{\text{RB}}] \leq \operatorname{Var}[\hat{f}_{\text{naive}}]$.

\noindent\textbf{2. Recursive Bayesian update of latent self-model.}

Given the conditional independence of \( X_t \) given \( \Phi^* \), the posterior over the latent self-model at time \( t \) follows a recursive update rule:
$P(\Phi^* | X_{1:t}) \propto P(\Phi^*) \prod_{i=1}^t P(X_i | \Phi^*)$.
This ensures that the posterior over \( \Phi^* \) evolves coherently across time, integrating past experiences into a stable, recursively updated self-model. Such updates are consistent with predictive coding and active inference frameworks.

\noindent\textbf{3. Emotional priors stabilize inference.}

By assumption, \( \Phi^* \sim P(\Phi^*) \) is affectively structured and temporally stable. This emotional structure guides inference by:
1) Assigning higher prior probability to self-consistent, goal-relevant hypotheses;
2) Filtering noisy observations and preventing overreaction to transient signals;
3) Regularizing belief updates to preserve alignment across context-content cycles.
Thus, Rao-Blackwellization using \( \Phi^* \) acts as a structure-preserving inference scheme that anchors episodic contents \( \Phi \) to a common latent frame.

\noindent\textbf{Conclusion.}
Together, these elements establish that inference conditioned on \( \Phi^* \) leads to:
\begin{itemize}
    \item Lower-variance estimates of cognitive-emotional content;
    \item Coherent, recursive updates of latent self-models;
    \item Stability and integration across temporally extended conscious episodes.
\end{itemize}
Hence, structure-preserving Rao–Blackwellized inference over \( \Phi^* \) formalizes how emotion-constrained Bayesian inference yields a coherent model of consciousness.

\end{proof}

\end{document}